# A bibliometric analysis on the current situation and hot trends of the impact of microplastics on soil based on CiteSpace

**Yiran Zheng[1], Yue Quan[1, 2]*, Su Yan[3], Xinting Lv[1], Yuguanmin Cao[2], Minjie Fu[1], Mingji Jin[1, 2]**

[1]Department of Resource Utilization and Plant Protection, Yanbian University, Yanji, 133002, China
[2]Department of Geography and Ocean Sciences, Yanbian University, Hunchun, 133300 , China
[3]Department of Agricultural Resources and Environment, Yanbian University, Yanji, 133002, China

E-mail: quanyue@ybu.edu.cn

## Abstract

In order to further understand the research status and development trends in the field of soil microplastics (MPs), this paper has collected and organized the research findings in the field of soil microplastics from the Web of Science Core Collection database from 2013 to 2024. Using the Web of Science Core Collection as the data source and with the assistance of the visualization software CiteSpace and VOSviewer, the literature related to the environmental impact of microplastics is taken as the analysis object. Through analyses such as keyword co - occurrence, keyword clustering, burst term analysis, author co - occurrence, and institutional co - occurrence of the literature related to the environmental impact of microplastics, we aim to better understand microplastics in the environment. microplastics accumulate in the soil, and then accumulate and transfer in organisms through the food chain, ultimately affecting the human body. Therefore, the analysis and detection of microplastics are of great significance for pollution prevention and control.This paper focuses on the research of the impact of international microplastics on soil and ecosystems, with the sample time from 2013 to 2024. The publication trend graph shows an upward trend in this area, peaking at 956 articles in 2024.There is a significant difference in the publication levels of authors, with a few highly productive authors achieving outstanding results. Keyword clustering has resulted in 10 major clusters, involving plastic pollution, microbial communities, etc. The research is divided into three stages: the preliminary exploration stage from 2013 to 2016, the expansion stage from 2017 to 2020, and the integration stage from 2021 to 2024. In the future, the research focus on the multi - level assessment of the impact of microplastics on soil ecosystems and their effects on soil organisms, in order to reveal the hazards and explore solutions.

## 1. Introduction

Plastic is a material that possesses several advantages over other materials, including its lightweight nature, wear resistance, low cost and stable chemical properties. Due to these benefits, plastic are widely used in various fields such as medical, clothing and industry [1]. In 2004, British scientist Thompson introduced the term 'microplastics' (MPs), which refers to plastic polymers with a particle size smaller than 5 mm. These microplastics are formed through the further degradation of used plastic waste via physical processes, photodegradation, and biodegradation. Common types of microplastics include polyethylene (PE), polypropylene (PP), polystyrene (PS), and polyethylene terephthalate (PET) [2]. Plastic products are ubiquitous in our daily lives and come into direct contact with environmental elements such as air, soil, and water.

The widespread application of microplastics, combined with their diminutive size and high specific surface area, enhances the adsorption of organic pollutants, heavy metals, and various environmental contaminants [3-5]. Furthermore, they act as intermediaries that link organisms to their surroundings, thereby influencing our living environment [6]. According to data from the European Plastics Association, global plastic waste production was estimated to range between 60 million tons and 99 million tons in 2015, with projections suggesting that this figure could potentially double or exceed by the year 2060 [7]. After the marine and freshwater environments, microplastic pollution of the soil environment is gradually gaining attention [8].

The sources of microplastics in soil are diverse. Firstly, the use of agricultural plastic mulch is an important source. According to the research by Astner, the global market scale of agricultural plastic film was 4 million tons in 2016, and it is expected to grow at a rate of 5.6% per year until 2030. The use of such a large amount of plastic film in agriculture undoubtedly poses a huge risk of microplastic pollution in farmland soil [9]. Secondly, surface runoff and agricultural irrigation are also sources. Even after being treated by a sewage treatment plant, a large number of microplastic particles still remain in the water [10]. According to research statistics, a small sewage treatment plant (with a treatment capacity of 390 $m^3$/h) releases $1.83\times10^{10}$ microplastic particles into the environment every day [11]. After the sewage is discharged into the water environment, it enters the soil through surface runoff and agricultural irrigation. At the same time, the application of manure is also an important source [9]. In the study by Blaesing [12], the average microplastic content in the tested organic fertilizer samples was 2.38 to 180 per kilogram. Finally, the sources of microplastics also include irresponsibly disposed plastic waste. The recycling rate of plastic products is less than 30%, and a large amount of waste plastic is disposed of by landfilling. Some of the plastic will further decompose into microplastics, which directly penetrate into the soil or enter the leachate. The effluent from the leachate treatment also contains a large amount of microplastics, which then enter the soil, thus posing a huge risk of soil microplastic pollution [13]. The fragmentation and pyrolysis of discarded plastic products, the use of sludge in agriculture, the application of organic fertilizers, agricultural irrigation, tire wear, and atmospheric deposition are all pathways through which microplastics enter the soil environment [14,15]. Microplastic pollution in soil and the soil ecosystem can affect the biogeochemical cycling processes on Earth. Microplastic pollution can also influence the dynamics of soil microbial communities by changing the physical and chemical parameters of the soil and the soil organic carbon content. In addition, microplastic pollution in the soil can trigger various toxic effects and increase the soil organic matter content, thereby affecting the natural microbial decomposition process [16]. Studies have shown that soil properties, including water-holding capacity, hydraulic conductivity, pore size distribution, and soil aggregation, are also affected by microplastic pollution [17]. Due to their strong hydrophobicity, large specific surface area, and abundant microporous structure, soil microplastics have a strong adsorption capacity for pollutants， which tend to adsorb organic and inorganic pollutants in the environment and aggregate with them to form new pollutant complexes, causing serious harm to the environment [18].

As carriers, microplastics cause the migration and transformation of pollutants, leading to persistent pollution with more serious consequences. After entering the soil, microplastics may have an impact on the physical and chemical properties of the soil, such as bulk density, permeability, water-holding capacity, and the structure of water-stable aggregates, thus reducing soil fertility. Existing studies have indicated that the presence of microplastics can be generally detected in biological tissues [19]. microplastics possess the ability to adsorb organic pollutants, heavy metals, and pathogens from their surroundings, leading to enhanced toxicological effects upon entering living organisms due to their interactions [20]. Extensive research has been conducted on the toxicological effects of microplastics on various organisms. Abidli S assessed the potential toxic impacts of polyethylene microplastics (PE-MPs) on Mediterranean mussels, revealing that low concentrations induced oxidative damage while high concentrations compromised the antioxidant system in bisexual digestive glands [21]. Another scholar, Ouyang, exposed common carp to environmentally relevant concentrations of microplastics for 30 days, followed by a 30-day period during which microplastics were excreted [22]. The researchers examined growth patterns, isotope and elemental composition, as well as intestinal microbiota, concluding that both intake and excretion of microplastics did not significantly impact isotope and elemental composition; however, they altered the structure and composition of the gut microbial community while diminishing functional diversity. MPs/NPs also exert substantial adverse effects on soil fauna, particularly earthworms and nematodes, influencing their growth, reproduction and lifespan through diverse toxic mechanisms, including bioaccumulation, DNA damage, genotoxicity, intestinal microbiota dysbiosis, histopathological damage, metabolic disorders, neurotoxicity, oxidative stress and reproductive toxicity. All these have an adverse impact on the natural ecological activities of these organisms, such as influencing litter decomposition, nutrient cycling and energy flow, causing various potential environmental and health hazards [23]. Furthermore, owing to their high surface area-to-volume ratio and hydrophobicity, MPs/NPs might act as transporters of pathogens and organic pollutants on land, transferring from the soil to plants and eventually to other organisms through the food chain, thereby constituting a threat to the environment [24]. Particularly in farmland soil, microplastics could be a more significant sink than that in ocean environment. The year-on-year accumulation of microplastics in farmland soil show crucial influences on the farmland soil ecosystem and thereby potentially pose a latent threat to the safety of food production and quality [25-27].

Hence, microplastic pollution in agricultural soil, as a novel environmental issue, has gradually gained attention and become a current research hotspot. The aforementioned influences of microplastics on the agricultural soil ecosystem are not merely caused by microplastics themselves; they can also form complex pollution with the pollutants existing in agricultural soil, thereby generating higher risks for the agricultural soil ecosystem.

To sum up, the aforementioned studies mainly concentrated on the integrated analysis of soil, water environment, crops, and microplastics. As microplastic pollution can cause severe harm, scholars both at home and abroad have paid extraordinary attention to this domain. Although there is a certain research foundation in the field of soil microplastics, it is still in the initial stage on the whole and lacks systematic integration and analysis. Therefore, in this paper, through the method of bibliometrics and with the aid of the CiteSpace visualization analysis software, a quantitative analysis was conducted on the titles, abstracts, authors, and publishing institutions of academic papers published in the field of soil microplastics. The purpose is to systematically and comprehensively analyze the current research volume, research teams, citation frequency, and predict future research trends in the field of soil microplastics, providing references for the further control of microplastic pollution in the soil environment and valuable references and innovative perspectives for researchers. With the aim of providing theoretical basis for the research breakthrough and comprehensive governance in the field of soil microplastics.

## 2. Methology

### *2.1 Database sources and search strategies*

The World's Top Academic Journal Database, Web of Science (WoS), is widely regarded as the most comprehensive and influential scholarly database across various fields, and has become a reliable resource for bibliometric analysis. Therefore, we utilized WoS to assess the current research landscape and identify the relationship between microplastics and soil. Due to the continuous evolution of the WoS literature database and the first publication of papers containing both "microplastics" and "soil" keywords dating back to 2013, we limited our analysis to the period from 2013 to 2024. This time period marks the beginning of our initial data collection phase. For "the impact of microplastics on soil," we conducted an advanced search in the WoS database: TS=(microplastics OR microplastic) AND (soils OR soil OR ground). Select "article" or "review" document type. By manually screening and deduplicating, we finally obtained 2,939 articles that elucidate the relationship between microplastics and soil. We exported the 2,939 articles from the WoS database in plain text format and processed them using the "Full Record and Cited References" as the source data for the bibliographic record.

### *2.2 Methods of analysis*

*2.2.1 Software selection.* This research utilized CiteSpace and VOSviewer software for the visualization of the knowledge graph, with the intention of enhancing the interpretability and readability of the relationships among the literatures. CiteSpace, an analytical tool developed on the basis of Java, has the ability to systematically extract crucial data, including publication information of literatures, journal distribution, research institutions, and the countries to which they belong. It then transforms this data into an associated network, and the influence of the literatures is intuitively represented through the size and color intensity of the nodes. By leveraging visualization technology, this software conducts a systematic description of knowledge resources and their carriers. It enables in-depth exploration, analysis, construction, and mapping of the knowledge system and its internal connections. In the form of a knowledge graph, it clearly reflects the development course, current research situation, hot issues, and future trend directions of the research field.

In the present study, through the creation of timeline diagrams depicting the year distribution and co-cited literatures, the annual research progress in the field of soil microplastics was systematically collated. By means of analyzing the collaborative networks of key classification nodes such as "authors", "institutions", and "countries", the cooperation models and the degree of cooperation within this field were investigated. Keyword co-occurrence analysis and clustering analysis methods were applied to uncover research hotspots, and subsequently predict the future development trends.VOSviewer possesses remarkable advantages in the clustering analysis of literatures. It can precisely expound the overall characteristics of the research field, efficiently depict and visualize the network structure of keywords, and quantify the correlations among the units within the network at the same time. In this research, VOSviewer and CiteSpace software complement each other in functionality, further refining the visual representation of research partnerships and the intensity of their cooperation. Consequently, it offers a multi-dimensional and systematic analytical viewpoint for the research in the field of soil microplastics.

*2.2.2 Methods of analysis.* A large number of literature data retrieved from the WoS database needs to be quickly identified and preprocessed, which is crucial for identifying the frontier and emerging trends in the field of research on the impact of microplastics on soil, and significantly affects the final results of the analysis based on CiteSpace and VOSviewer. Therefore, we first downloaded the complete records and citation data from WoS and saved them as "download_XXX.txt" files. Then, we imported the data into the CiteSpace software and set the time slice to 1 year, specifying "node type" as "keyword" and using other parameters for co-citation network analysis. We chose "pruning slice network" as the connection pruning method and used the "show merged network" function to display the comprehensive analysis chart of the filtered data. By setting "node type" to "keyword" and analyzing the co-citation network, we used "show merged network" to present the complete analysis chart of the filtered data. Similarly, after importing the literature data in WoS format into the Java environment, each node in the visualization network represents a journal and an author, etc. Keyword. Node size represents the occurrence frequency, while node color represents the analyzed keyword and clustering. In the CiteSpace analysis tool, the centrality of the node reflects its critical role in the network structure and reflects the degree of connection with other nodes. This indicator is used to quantify the importance of the node in the network. In simple terms, the higher the node centrality value, the more important the keyword is in the co-occurrence network. Finally, the partnerships among countries were visualized using ScimagoGraphica software (figure 1).

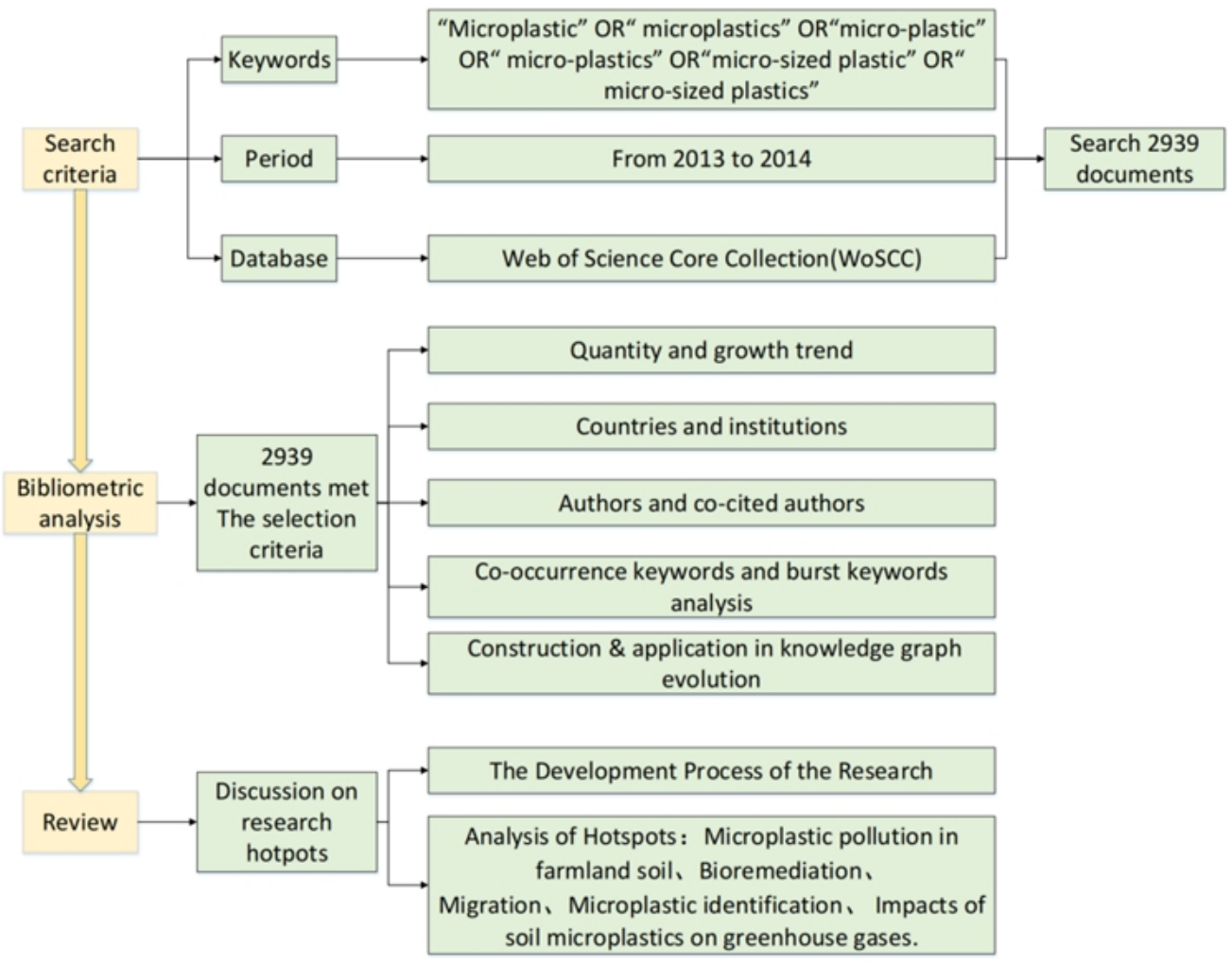

Figure.1. Paper selection and flow chart of research framework

## 3. Results

### *3.1 Total number of publications*

The number of publications in the field of soil microplastics is shown in figure 2. The data shows the number of publications in the field of soil microplastics. Among them, the data obtained from WOS shows that the research on soil microplastics has gone through a slow initial stage and a rapid development stage, with the number of publications increasing sharply. It can be seen that the first research paper appeared in 2013; during the initial stage from 2014 to 2016, only one paper was published; from 2016 to 2017, 8 to 16 papers were published, which can be regarded as the slow initial stage.From 2018 to 2022, the number of published papers in the field of soil microplastics has increased exponentially, even reaching nearly 600 in 2022. The research on soil microplastics has entered the initial stage of rapid growth, indicating that both domestic and foreign scholars are continuously paying attention to the field of soil microplastic pollution. At the same time, the continuous high-speed growth for several years also indicates that there are still many problems to be explored, with broad research prospects.

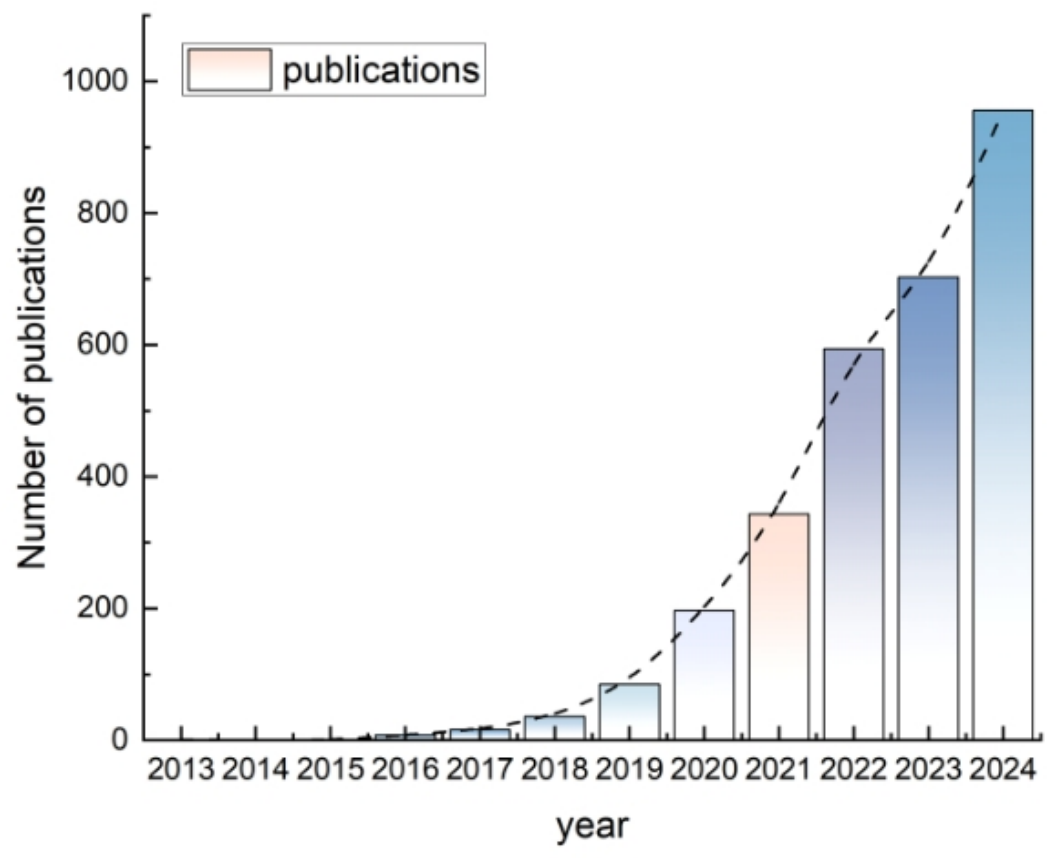

Figure. 2. Quantity of publications on microplastic in soil environmental field

### *3.2 Scientific collaboration analysis*

To obtain information on the number of published articles on soil microplastics research from various countries around the world, this paper used the Citespace and VOSviewer softwares (figure 3 and figure 4) to extract the information on the countries and regions where the articles were published. From figure 3 and figure 4, it can be seen that from 2013 to 2024, a total of 187 institutions from 103 countries participated in this research. In figure 3, judging by the size and color of the nodes, the node of " CHINA" is the largest and has a brighter color, indicating that China has published more results and is more active in this field of research, and is the main force in the research. The nodes of the United States, India, and other countries are also relatively large, suggesting that they also have high research outputs in this field. There are connections between China and several countries such as Germany, the United States, and Netherlands, indicating that China has carried out cooperative research with many countries in this field. There are also close cooperative relationships among European countries, such as Germany and Switzerland, Italy, etc. Moreover, although the nodes of Brazil, Saudi Arabia, etc. are relatively small, they have also established certain cooperative connections with other countries through connections. The time color spectrum in the lower left corner shows that different colors correspond to different years. It can be seen that the research in this field of various countries has gradually developed since 2013, and the research enthusiasm has continued to rise in recent years (the warm tone part), and the cooperative relationships have also been continuously expanded and deepened. Countries with more cooperation may continue to deepen cooperation and jointly tackle key research directions. At the same time, some countries that currently have less cooperation, such as some countries in Asia and South America, may strengthen exchanges and cooperation with each other in the future and expand the international network of research. In addition to the currently active countries, some emerging countries may increase investment in this field, enhance research capabilities and outputs, and the nodes in the graph will gradually increase and become new research forces. China published the most articles on soil microplastics research, a total of 1626. This is because China has always attached great importance to soil quality, the health and lives of the people, especially in recent years, the Chinese government has successively issued the "Action Plan for Soil Pollution Prevention and Control", "Soil Environmental Quality Standards for Agricultural Land Contamination Risk Control (for Trial Implementation)" (GB 15618-2018) and other laws and regulations, and has carried out a nationwide soil quality survey and built high-standard farmland throughout the country. On the one hand, with the rapid development of science and technology and the economy, China is playing an increasingly important leading role in various global environmental problems. On the other hand, the Chinese government attaches great importance to international cooperation and exchange in soil microplastics research field, actively participates in international academic conferences and symposia on soil microplastics research, and conducts extensive cooperation and exchange with international peers. Concerned about the problem of new pollutants, and actively encourages universities and research institutions to focus on research in the field of soil microplastics (for example, with regard to plastic film (one of the main sources of soil microplastics), China is a large country, and the area covered by agricultural plastic film in the use of land is the first in the world. If a country has a high demand for plastic film, then the research investment in this area will also increase accordingly, so the number of research results will naturally increase. The United States ranks second with 289 articles. The United States leads the world in environmental technology. Other countries such as the Netherlands, Germany, the United Kingdom, and Australia, which have published more articles in the field of soil microplastics (table 1), are also countries that have received Additionally, the Chinese Academy of Sciences (CAS) is the largest contributor to date (figure 4) with the highest centrality (0.93). This result indicates that the Chinese Academy of Sciences has extensive contacts with other institutions. Although 218 authors participated in research in this field, more than 80% of the authors had no connections to each other (centrality value of zero) (figure 5). These results suggest that more cooperation and communication are needed.

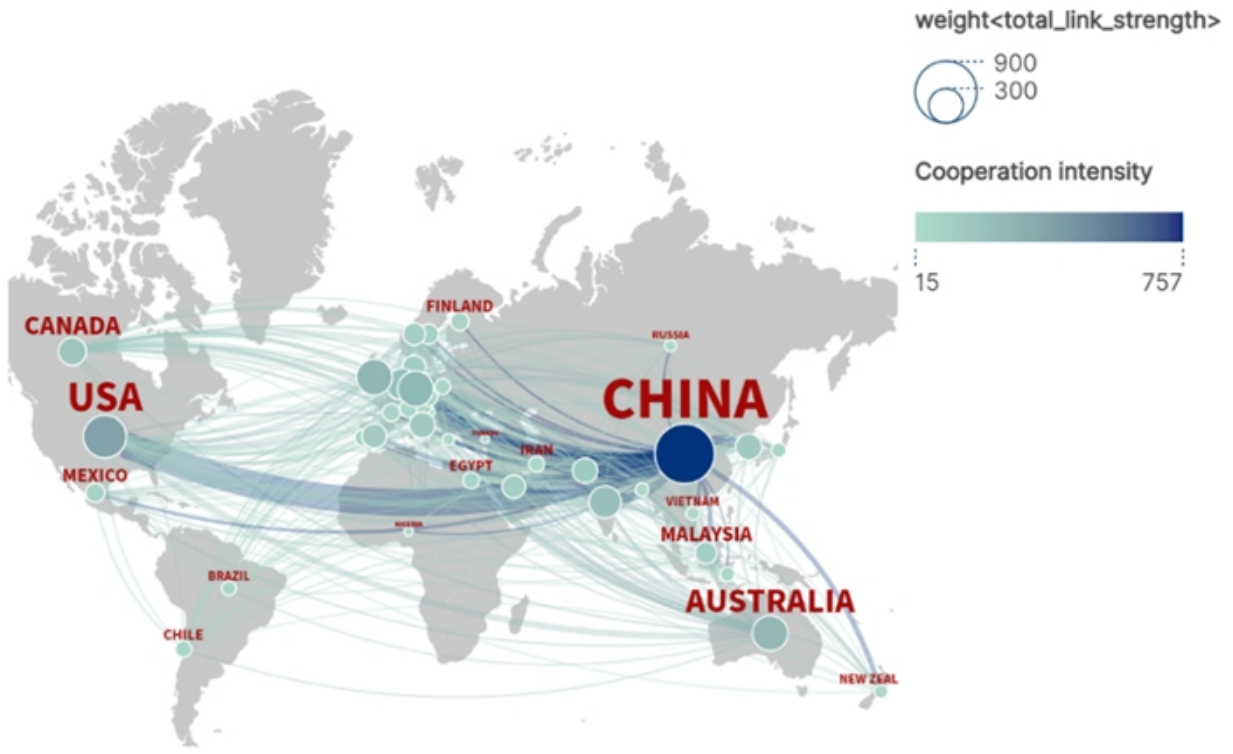

Figure.3 .Cooccurrence map of source countries in soil MPs field in WOS (Citespace software uses the diameter of the circle to indicate the amount of posts in the result presentation. The larger the diameter, the m

Table 1. The count and centrality in different countries.

| Count | Centrality | Year (First Appear) | Countries/Regions |
|---|---|---|---|
| 1626 | 0.08 | 2017 | CHINA |
| 289 | 0.18 | 2017 | USA |
| 224 | 0 | 2016 | GERMANY |
| 202 | 0 | 2019 | INDIA |
| 169 | 0.12 | 2016 | AUSTRALIA |
| 135 | 0.41 | 2013 | ITALY |
| 130 | 0.04 | 2018 | SOUTH KOREA |
| 111 | 0 | 2016 | NETHERLANDS |
| 109 | 0.37 | 2013 | ENGLAND |

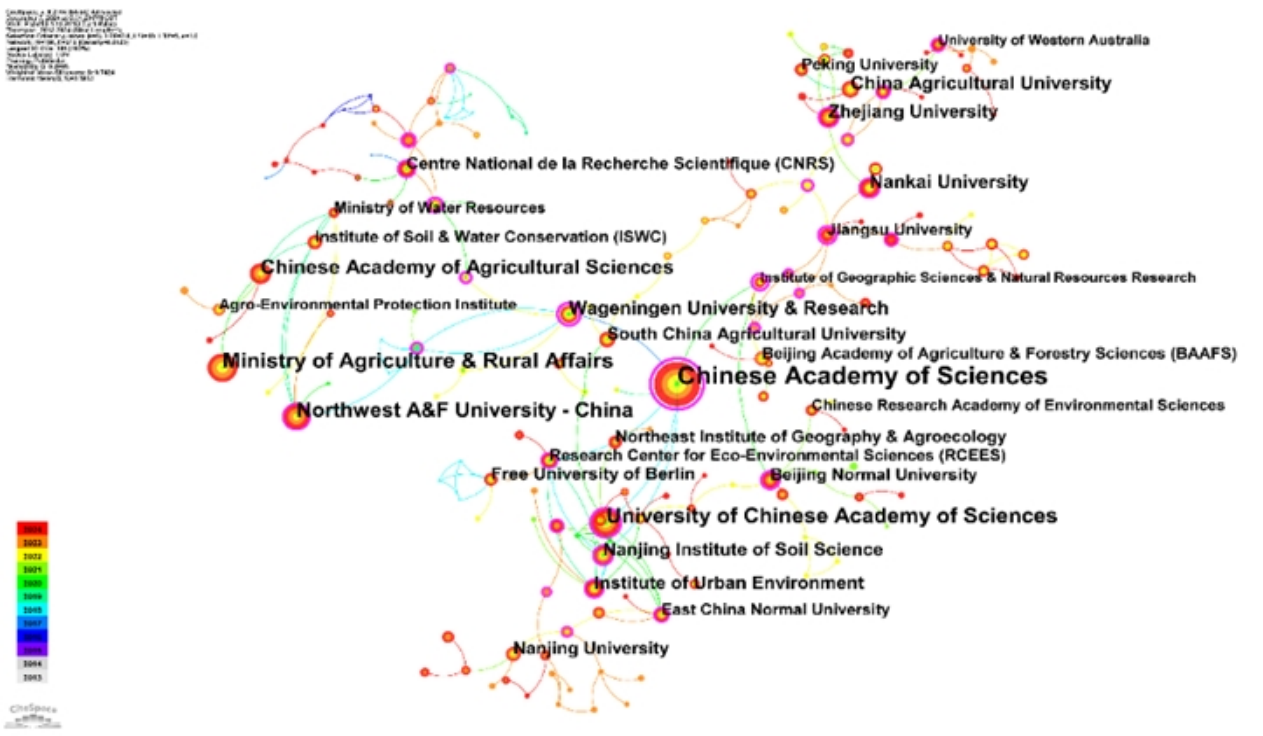

Figure. 4. Cooperation networks for institutions that have performed research on soil MPs. The different color of the line between the nodes represents the time when the relevant institutions appeared, where dark colors

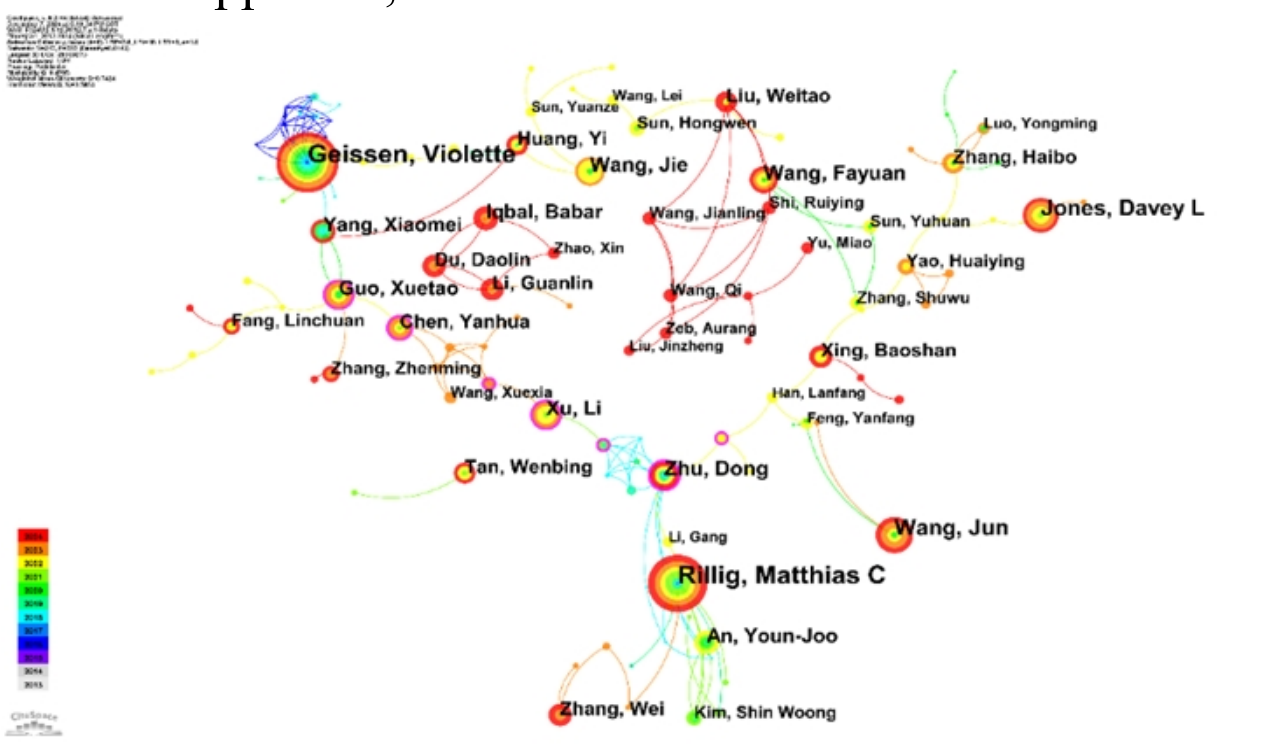

Figure. 5. Author cooperation network in the field of soil MPs. The different color of the line between the nodes represents the time when the relevant institutions appeared, where dark colors represent earlier years

### *3.3 Citation analysis*

Overall, the main relevant publications were concentrated from 2017 to 2018. As can be seen from table 2, this period was the initial stage of the field of soil microplastics, laying the foundation for the rapid growth of subsequent publications. Among them, the paper "Microplastics in freshwater and terrestrial environments: Evaluating the current understanding to identify the knowledge gaps and future research priorities" published by Horton A A in 2017 had the highest citation frequency from 2000 to 2023 [28]. The team believed that freshwater and terrestrial environments are the sources and pathways for plastics to flow into the ocean, and microplastics tend to accumulate on land, especially in areas with intensive human activities such as urban and agricultural areas. For example, a study on the farmland soil in the suburbs of Shanghai, China, found that the farmland near industrial areas had a high degree of microplastic pollution, which originated from industrial emissions, atmospheric deposition, and surface runoff, revealing the profound impact of urbanization on the distribution of soil microplastics. The study on soil aggregates in southwestern China showed that the distribution of microplastics was uneven under different land use types, and the abundance of microplastics in farmland soil was relatively high, which was closely related to the use of agricultural films and sewage irrigation. The second most cited paper was "Microplastics Can Change Soil Properties and Affect Plant Performance" published by the team of Machado, AAD in 2019 [29]. The third most cited paper was "Impacts of microplastics on the Soil Biophysical Environment" published by the team of Machado, AAD in 2018 [30]. Both papers focused on the impacts of microplastics in the soil environment. The core of both was to explore how microplastics act on the soil and related components of the ecosystem, aiming to reveal the potential effects of microplastics on the soil ecosystem and provide a scientific basis for understanding the ecological consequences of microplastic pollution. Both papers emphasized the soil ecosystem. The properties of the soil itself, such as physical and chemical properties, and the biological components in the soil, such as plants, soil animals, and microbial communities, were important research objects in both papers, highlighting the comprehensive impact of microplastics on the soil - biological system. However, the former focused more on the direct changes in soil properties caused by microplastics and how these changes further affected the growth performance of plants. It elaborated on the effects of microplastics on the physical properties of the soil (such as pore structure, air permeability, and water - holding capacity) and chemical properties (such as nutrient adsorption and additive release), and then explored how these changes affected the germination rate, plant height, biomass, and physiological processes such as photosynthesis and respiration of plants, with the growth performance of plants as an important research focus. The latter paid more comprehensive attention to the various impacts of microplastics on the soil biophysical environment. In addition to the impacts on soil physical properties (such as the influence of soil structure and porosity on soil aeration and water movement) and the effects on soil biological communities (microorganisms and soil animals), it also emphasized the interactions between microplastics and other substances in the soil (such as nutrients and heavy metals). Relatively speaking, the research scope was broader, covering more interactions among biophysical factors in the soil ecosystem.

Table 2. Top 10 Articles on WoS Relating to Published Research on the Impact of MPs on Soil

| Rank | Title | The first author | Citation frequency | Publishing time |
|---|---|---|---|---|
| 1 | Microplastics in freshwater and terrestrial environments:Evaluating the current understanding to identify the knowledge gaps and future research priorities [28] | Horton A A | 2375 | 2017 |
| 2 | Microplastics Can Change Soil Properties and Affect Plant Performance [29] | Machado, AAD | 1130 | 2019 |
| 3 | Impacts of Microplastics on the Soil Biophysical Environment [30] | Machado, AAD; | 1039 | 2018 |
| 4 | The distribution of microplastics in soil aggregate fractions in southwestern China [31] | Zhang, GS | 944 | 2018 |
| 5 | Microplastic and mesoplastic pollution in farmland soils in suburbs of Shanghai, China [32] | Liu, MT | 912 | 2018 |
| 6 | Microplastics in Swiss Floodplain Soils [33] | Scheurer, M | 877 | 2018 |
| 7 | Effects of Microplastics in Soil Ecosystems: Above and Below Ground [34] | Boots, B | 856 | 2019 |
| 8 | Macro- and micro- plastics in soil-plant system: Effects of plastic mulch film residues on wheat (Triticum aestivum) growth [35] | Qi, YL | 794 | 2018 |
| 9 | Source, migration and toxicology of microplastics in soil [36] | Guo, JJ | 708 | 2020 |
| 10 | Microplastics in soils: Analytical methods, pollution characteristics and ecological risks [37] | He, DF | 668 | 2018 |

The data provided by the sample research was processed and analyzed, and the project was selected as "author". A co-authorship map related to the international research on the impact of microplastics on soil and ecosystems was drawn, as shown in figure 6. It can be seen from the figure that the number of connections is greater than the number of nodes, and the density is relatively high. This indicates that there are many cooperative relationships among researchers who have made significant contributions to this field internationally, and they tend to publish papers in collaboration with other institutions and groups rather than independently. It is not difficult to see that several authors, led by Geissen, Violette and Rillig, Matthias C, have close cooperation and have formed a certain scale. Their nodes are large, indicating that they have published more research results and may have been cited more frequently, thus having a high influence in the field of microplastics research. The connections between nodes represent the cooperative relationships among scholars. Among them, Geissen, V mainly focuses on the traceability of soil microplastics [38-40], the impact of soil microplastics on microorganisms and animals and plants [41-43], as well as the identification methods of microplastics in soil [44,45], and has established cooperative relationships with the high-yield author Yang, XM (16 articles) in China. It can be seen that "Geissen, Violette" has connections with many scholars such as "Yang, Xiaomei" and "Guo, Xuetao", indicating that his cooperation is extensive. "Rillig, Matthias C" also has cooperative relationships with scholars such as "An, Youn-Joo" and "Kim, Shin Woong". In addition, there are also cooperative connections among Chinese scholars with the surname "Wang", forming a certain cooperative group. It can be seen from the figure that the related research has gradually developed since 2013, and has remained active in recent years (the warmer color areas), with new scholars constantly joining the research and the cooperative network continuously expanding.

To ensure the accuracy of the data and avoid the situation of cross-statistics of articles, only the first author of the literature was retained when organizing and statistics were made in this paper. After software operation, it was statistically found that there were 217 authors who published articles, among which 46 authors published 7 or more articles, 14 authors published 6 articles, 24 authors published 5 articles, 38 authors published 4 articles, 13 authors published 3 articles, 22 authors published 2 articles, and 60 authors published 1 article.

From the above table 3, it can be seen that there are many authors conducting research on the impact of international microplastics on soil and ecosystems, and the research team is relatively large. However, the publication levels and research intensities of these authors vary greatly. There are 46 authors with 7 or more publications, accounting for 18.89% of the total number of authors, and their publication volume is 573, accounting for 51.76% of the total publication volume. Compared to others, this group is relatively small but has published a large number of research papers, indicating that they have strong productivity and numerous research achievements in the field of international microplastics ' impact on soil and ecosystems. There are 95 authors with 3 or fewer publications, accounting for 43.78% of the total number of authors, suggesting that a large number of researchers have only 1 to 3 research results. Based on the above analysis, it can be found that the publication levels of authors conducting research on the impact of microplastics on soil and ecosystems vary greatly. Only a few researchers have conducted in-depth and multi-faceted studies on the impact of microplastics on soil and ecosystems, while the majority of researchers have only conducted single studies on a certain aspect and have not conducted systematic and in-depth research or multi-faceted studies, indicating insufficient investment in this field.

From the overall map, it can be seen that the connections between scholars from different countries are rich, showing a trend of international cooperation. This international cooperation has broken geographical restrictions and brought diverse research ideas and methods. In terms of the time dimension, before 2020, during the initial stage, foreign researchers and research teams dominated, and the cooperation was relatively loose. However, after 2020, the number of Chinese researchers and research teams has rapidly increased, and the cooperation has become more closely-knit.

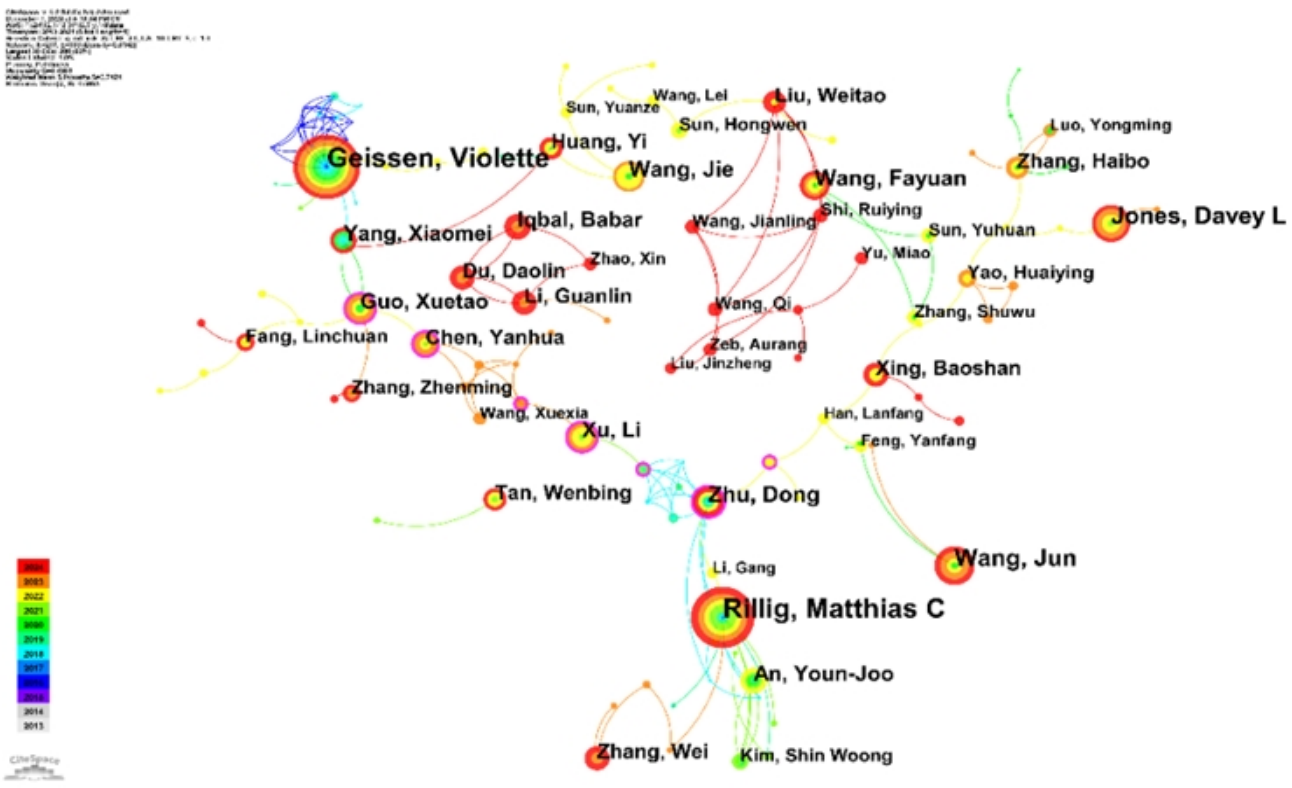

Figure. 6. Cooperation network of authors in the field of MPs pollution in soil

Table 3. The number of authors with different amounts of publications

| Number of published articles | Number of authors | Percentage of authors out of the total number | Percentage of the number of published papers in the total publications |
|---|---|---|---|
| 40 | 1 | 0.46% | 3.61% |
| 38 | 1 | 0.46% | 3.43% |
| 24 | 1 | 0.46% | 2.17% |
| 23 | 1 | 0.46% | 2.08% |
| 19 | 1 | 0.46% | 1.72% |
| 18 | 2 | 0.92% | 3.25% |
| 17 | 2 | 0.92% | 3.07% |
| 16 | 3 | 1.38% | 4.34% |
| 15 | 4 | 1.84% | 5.42% |
| 14 | 5 | 2.30% | 6.32% |
| 11 | 3 | 1.38% | 2.98% |
| 10 | 2 | 0.92% | 1.81% |
| 9 | 8 | 3.69% | 6.50% |
| 8 | 7 | 3.23% | 5.06% |
| 7 | 5 | 2.30% | 3.16% |
| 6 | 14 | 6.45% | 7.59% |
| 5 | 24 | 11.06% | 10.84% |
| 4 | 38 | 17.51% | 13.73% |
| 3 | 13 | 5.99% | 3.52% |
| 2 | 22 | 10.14% | 3.97% |
| 1 | 60 | 27.65% | 5.42% |

### *3.4 Analysis for Keyword Co-Occurrence*

As illustrated in figure 7, the network visualization is constituted by assorted nodes and interconnecting wires. These keywords are categorized into eight distinct clusters. In this graphical representation, the magnitude of the nodes mirrors the frequency of the corresponding keywords, while the coloration of the nodes designates the cluster to which they pertain, with each cluster being demarcated by a unique color.

Within the green cluster, "microplastics" is positioned at the heart of the green area and constitutes the central focus of the overall study. Encircling it are terms such as "nanoplastics", "heavy metals", and "ecosystem" that are interconnected. This suggests that the research within this domain pivots around the intrinsic characteristics of microplastics and their reciprocal interactions with pollutants like heavy metals within the ecosystemic context. Moreover, it encompasses the cascade of consequences that microplastics precipitate at the ecosystem level, notably the perturbations on the ecosystem's structural and functional aspects.The blue cluster is anchored by "soil" and "degradation" as its nuclei and is flanked by lexical items like "agricultural soil", "wastewater", "water treatment plants", and "plastics". This configuration indicates that the research efforts in this realm concentrate on microplastics-related inquiries within the soil milieu. It encompasses the prevalence of microplastics in agricultural soils and their degradation kinetics during the treatment of wastewater and sewage, encapsulating the dynamic metamorphoses of microplastics in scenarios germane to soil ecology and sewage treatment procedures.In the red cluster, "toxicity" and "pollution" feature prominently and are allied with terms like "organic matter", "microbial community", "biodegradation", and "growth". This implies that the investigative spotlight here is trained on dissecting the toxic repercussions instigated by microplastic pollution and its ramifications for the microbial community and biological proliferation. Concomitantly, it assimilates aspects related to toxicity and pollution during the biodegradation chronicle of microplastics. The yellow cluster gravitates around "marine environment" and "water" as its epicenters and is tethered to terms such as "freshwater", "surface waters", "plastic debris", and "accumulation". Manifestly, this area predominantly zooms in on microplastic quandaries within aqueous media, with particular emphasis on the accretion of microplastics in oceanic and freshwater habitats and the comportment of plastic debris in water bodies, spotlighting the dynamic flux of microplastics across diverse aquatic settings.Within the purple cluster, "particles" and "environment" assume greater salience and are correlated with "fibers", "risk assessment", and "knowledge gaps". This signals that the research in this precinct revolves around the environmental footprint of microplastic particles, assimilating elements related to microplastic fibers and ruminations on the environmental risk appraisal of microplastics. Simultaneously, it remains attuned to the extant lacunae in current knowledge reservoirs. The turquoise cluster orbits around "ecological risk" and "removal" as its cardinal words and is affiliated with "sewage sludge", "ecological indicators", and "bioremediation". This intimates that this area is centered on the ecological perils posed by microplastics, probing into methodologies for microplastic abatement via bioremediation and leveraging ecological indicators to gauge the ecological risks stemming from microplastic pollution, with a focal point on microplastic pollution remediation and ecological risk containment.

Finally, in the orange cluster, “water” emerges as the linchpin keyword and is associated with “sorption”, “transport”, and “nanoparticles”. This area principally hones in on the behavioral patterns of microplastics within aqueous media, such as the adsorption kinetics of microplastics in water and research efforts on the migratory traits related to nanoparticles, unearthing the dynamic transformation mechanisms of microplastics in the water environment.

Keywords are a precise summary of the content of an article. By using CiteSpace to analyze keyword frequency and centrality, the top 10 hot keywords in the field of soil microplastics research (table 4) were obtained from the perspective of microplastics and soil degradation.

From the perspectives of microplastics and soil degradation, research has shown that plastic film coverings are an important source of microplastic pollution in soil due to their effective contribution to crop production [46]. Consequently, the global use of PE plastic film is increasing year by year . It is therefore crucial to develop biodegradable film alternatives to traditional plastic film and to evaluate their agricultural application effects and widely promote their use [47]. Small size, colored, fiber-like, and microspherical are the main characteristics of microplastics. The influence of different particle sizes and types of plastic particles on soil matrix varies. Yu H found that 95% of the sampled soil samples had plastic particles with a size of 0.05-1mm by investigating the abundance and distribution of plastic particles in the soil aggregates of four planting areas and an established riparian buffer strip in southwestern China. The main form was plastic fibers (92%), followed by fragments and thin films (8%) [27]. The concentration of plastic particles in the buffer strip soil was lower than that in the adjacent vegetable field soil, indicating that it is necessary to control the use of soil amendments and wastewater irrigation to reduce the accumulation of microplastics in agricultural soil. Research has shown that the abundance of microplastics in soil is not linearly related to soil pH, dissolved organic matter content, and iron content , but the increase in microplastic content will cause an increase in soil pH, ammonium nitrogen content, and dissolved organic carbon content and bring about a significant increase in the average abundance of some resistant degradable microorganisms [48]. Moreover, the additives used in plastic products may change the structure of soil microbial communities and lead to soil degradation , affecting the global carbon-nitrogen cycle and leaving more hidden dangers for the material and energy flow in soil [49].

From the perspective of pollution, the surface of microplastics deteriorates after aging, and their surface hydrophobicity decreases and hydrophilicity increases [37]. Therefore, when plastic microparticles reach land, their surfaces with functional modifications and large specific surface areas facilitate the adsorption of dissolved substances.

Table 4. The top 10 hot keywords in the field of soil MPs research

| Rank | Keywords | Frequency | Centralit |
|---|---|---|---|
| 1 | pollution | 668 | 0.08 |
| 2 | microplastics | 517 | 0.3 |
| 3 | soil | 434 | 0.22 |
| 4 | water | 328 | 0 |
| 5 | degradation | 302 | 0.17 |
| 6 | plastics | 296 | 0.16 |
| 7 | accumulation | 250 | 1.01 |
| 8 | marine environment | 247 | 0.69 |
| 9 | particles | 245 | 0.32 |
| 10 | transport | 229 | 0.72 |

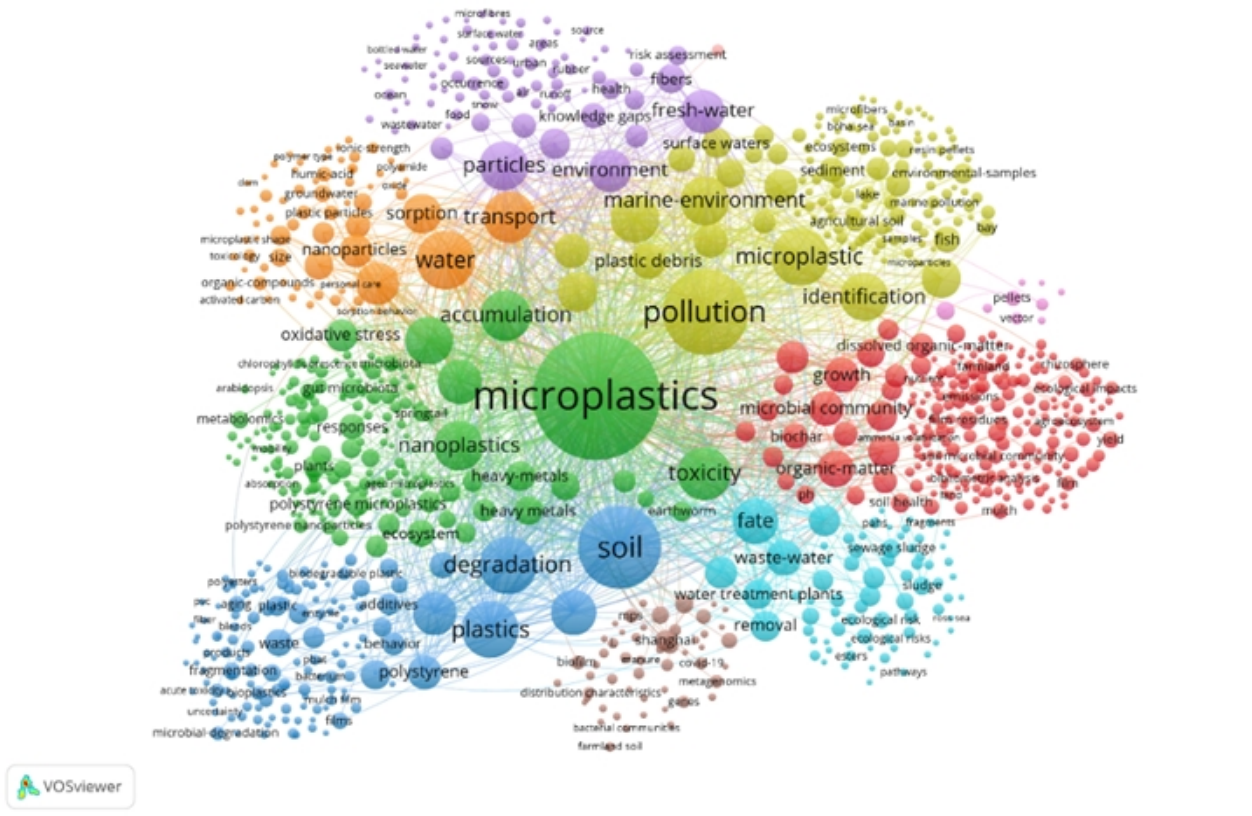

Figure. 7. The network visualization map of co-occurrence keywords for the field of MPs pollution in soil

### *3.5 Analysis of Keyword Clustering*

A clustering analysis of the keywords was conducted, and the result is shown in figure 8, with a total of 9 clusters of keywords. The main characteristic parameters of keyword clustering are two: (1) The clustering module value (Q value), and it is generally considered that Q > 0.3 indicates a significant clustering structure; (2) The S value: The average silhouette value of the clustering. S > 0.5 indicates that the clustering is reasonable, and S > 0.7 indicates that the clustering is convincing. The Q value of this clustering is 0.4279 > 0.3, which means that the clustering structure identified is significant. The S value is 0.695, very close to 0.7, and it can be considered that the result of the clustering is convincing.

The research imported the standardized data of the sample study into the software CiteSpace and drew a keyword clustering map related to the research on the impact of microplastics on soil and ecosystems abroad (figure 8). From the clustering map, it can be seen that there are a total of 10 major clusters, namely plastic pollution, microbial communities, surface waters, microbial community, oxidative stress, marine environment, soil properties, heavy metals, transport, risk assessment, and photosynthesis. Therefore, it is not difficult to find that international research on the impact of microplastics on soil and ecosystems is closely related to plastic pollution, microbial communities, surface water, etc.

There is a very complex interaction between microplastics and microbial communities. Studies have shown that microorganisms may use microplastics as a carbon source for degradation, but this degradation process is very slow and limited [50]. Microplastics may change the diversity of microbial communities, inhibit the activity of some beneficial microorganisms, and even promote the reproduction of some harmful microorganisms [51]. Microorganisms in the soil have multiple functions such as degrading organic pollutants, promoting plant growth, and improving soil structure. As the impact of microplastics on soil health becomes more severe, the use of microbial communities for soil remediation and pollution control has gradually attracted the attention of scholars. At the same time, the European Union's "European Green Deal" and plastic strategy also support the research on biodegradable plastics and microbial technologies for microplastic pollution control, further promoting the application of microorganisms in plastic pollution control. Therefore, microbial communities have become a research hotspot for scholars [52]. In the future, in order to improve the efficiency of microbial degradation of microplastics, scientists will need to develop new technologies or optimize existing microbial community cultivation and application methods to enhance their application potential in soil remediation.

The microbial communities in the soil are highly sensitive to the presence of microplastics [53]. Microplastics may change the population structure and metabolic activities of microorganisms, affecting soil nutrient cycling, carbon fixation and other ecological functions [54]. The potential toxicity of microplastics to microorganisms and their impact on microbial diversity may further weaken the self-repair and maintenance capacity of the soil. Many countries have mentioned in their laws and policies on plastic pollution that soil health and agricultural productivity may be negatively affected by microplastic pollution [55]. Therefore, governments around the world have also begun to promote soil research related to microplastic pollution and invest funds to support related projects [56]. For example, environmental protection agencies in the United States, Australia, and Japan have funded research on the relationship between microplastics and changes in soil properties. With the rapid development of modern soil science, environmental science, and ecology, scientists can more accurately detect the types, concentrations, and impacts of microplastics in the soil, thereby optimizing agricultural production systems, improving soil health and productivity, and promoting sustainable agriculture. Therefore, soil properties have become a research hotspot for scholars.

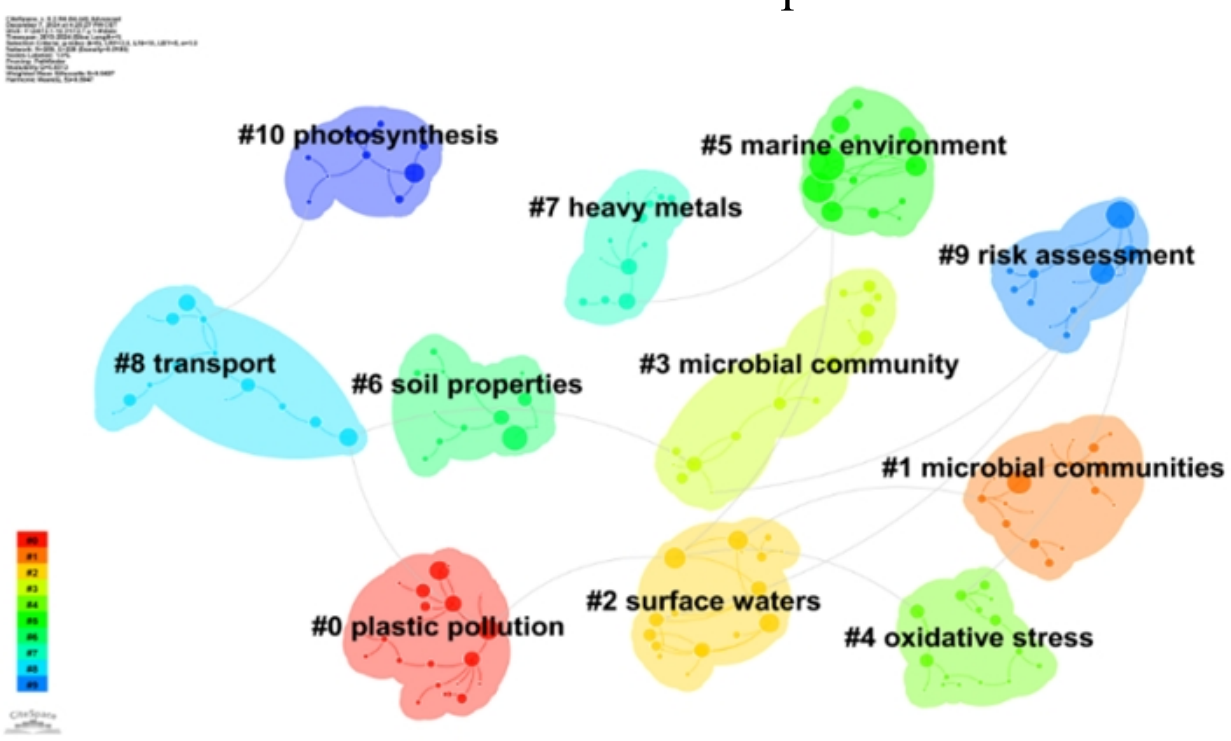

Figure. 8. Keyword clustering map of MPs in soil

### 3.6 Timeline

The timeline chart, often referred to as the "Timeline," is constructed by a sequential arrangement of a series of horizontal time bands along the chronological axis. It primarily serves to depict a knowledge graph illustrating the evolution of keywords through the temporal dimension. In the context of this study, the sample data underwent meticulous processing before being imported into the CiteSpace software. Subsequently, the "keywords" were chosen as the focal point for analysis within the project framework. A keyword co-occurrence map was then meticulously crafted, encapsulating the research and applications concerning the impact of microplastics on soil and ecosystems. Building upon this map, a comprehensive analysis of the keyword timeline was conducted, leading to the generation of a timeline chart that specifically delineates the research trajectory related to the influence of microplastics on soil and ecosystems (figure 9).

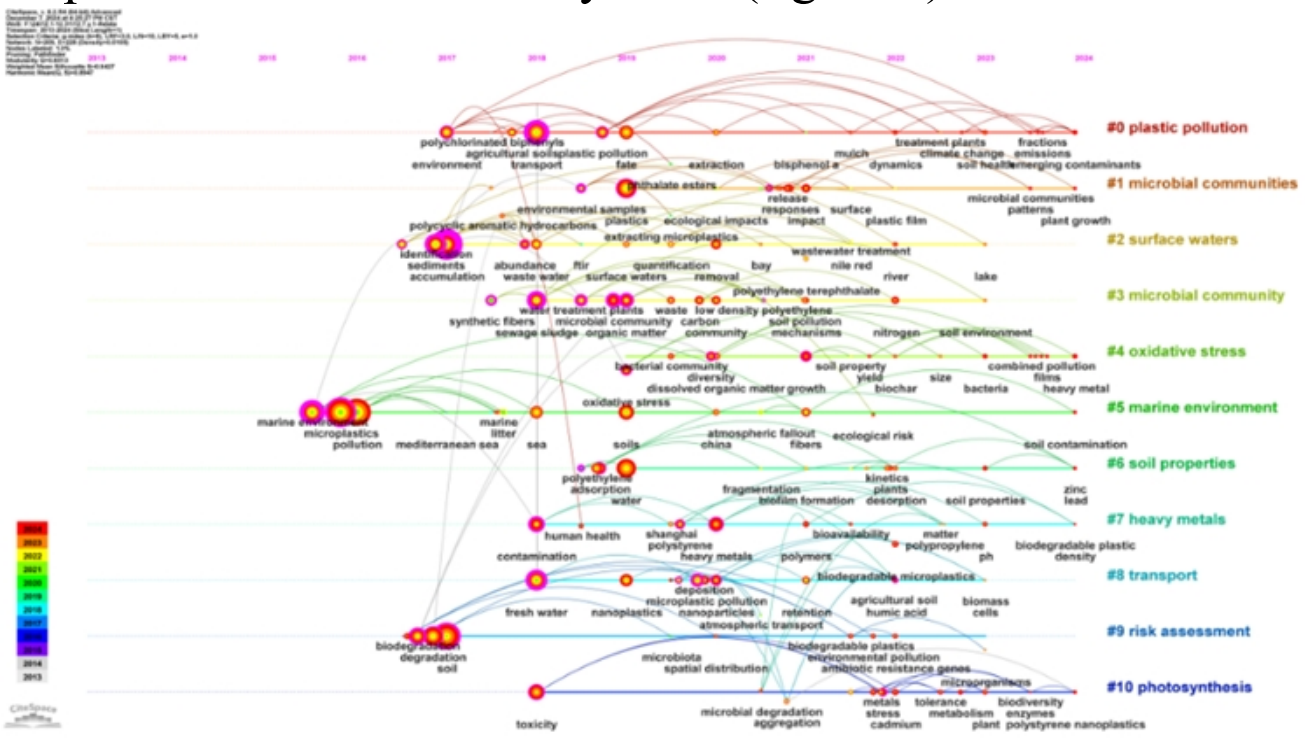

Figure. 9. Keyword clustering timeline map

The timeline chart offers a valuable vantage point for discerning the developmental trends and cutting-edge frontiers of research into the impact of microplastics on soil and ecosystems from a temporal perspective. The keywords are arranged in a left-to-right sequence, mirroring the chronological progression of their changes. In the accompanying figure, the size of each node is directly proportional to the frequency of the corresponding keyword's appearance; larger nodes signify a higher frequency of occurrence.

#2 Keywords related to “surface waters”, such as “river” and “lake”, are of significance. Surface water serves as a crucial water resource for humans and an important reservoir for pollutants. The research centers on its pollution sources, migration, transformation and ecological impacts. Surface water bodies constitute an integral and pivotal component of the ecosystem, and fluctuations in water quality can have far-reaching implications for the equilibrium of the entire ecological system. As public awareness of water pollution intensifies, scholars have increasingly recognized that the ramifications of water contamination extend beyond aquatic organisms, potentially permeating soil, plants, and terrestrial ecosystems via the hydrological cycle. Moreover, the microbial community structure within water bodies serves as a sensitive indicator of water quality variations. In this regard, the concepts of "microbial community" and "surface water" are mutually reinforcing, and they have concurrently become prominent research topics among scholars [57]. #6 Soil properties: The research on microplastics in soil is also quite prominent. The use of plastic films in agricultural production, the agricultural application of sludge, and atmospheric deposition are the main sources of soil microplastics. Microplastics can affect the physical properties of soil (such as porosity and water permeability), chemical properties (such as nutrient cycling and pollutant adsorption and desorption), and soil biota (such as microorganisms and soil animals). For example, microplastics may change the community structure and function of soil microorganisms, affecting the decomposition of soil organic matter and nutrient transformation.#0 plastic pollution): As the core theme, it covers all aspects of microplastic pollution, including pollution sources (such as the use of disposable plastic products and poor plastic waste management), the current pollution situation (surveys of pollution levels in different regions around the world), and pollution control strategies (such as the formulation of policies and regulations and source reduction measures).(#9 risk assessment): The risk assessment of microplastics to ecosystems and human health is an important research direction. Ecological risk assessment involves evaluating the degree of harm of microplastics to different biological groups (from microorganisms to higher organisms) and ecosystem functions; human health risk assessment focuses on the potential hazards that microplastics may cause to human organs and physiological functions after entering the human body through food chains, respiration, and other pathways. Findings from studies conducted by the United Nations Environment Programme and other reputable institutions have revealed that plastic waste has emerged as one of the most significant contributors to marine pollution [58]. Each year, millions of tons of plastic waste infiltrate the ocean, posing a grave threat to marine life as fish, birds, and marine mammals frequently ingest plastic debris, which can be lethal. In light of this pressing issue, there is an urgent global imperative to explore alternative solutions and expedite the research and implementation of degradable plastics [59]. Presently, an increasing number of countries and regions have implemented measures to prohibit or curtail the production and consumption of disposable plastic products. A notable example is the European Union's adoption of the "Single-Use Plastics Directive" in 2019, which aims to reduce the reliance on disposable plastic items and stimulate the development of biodegradable alternatives. Early generations of biodegradable plastics were plagued by issues such as inconsistent degradation rates. To address these challenges, it is imperative to leverage technological innovation to enhance the strength, flexibility, and longevity of degradable plastics. Looking ahead, future generations of degradable plastic materials should be engineered to adapt their degradation rates in response to varying environmental conditions, including temperature, humidity, and light exposure. This would endow degradable plastics with greater versatility across a diverse range of applications, such as food packaging, medical supplies, and agricultural films.

By interpreting this map, we can clearly see that rich achievements have been made in microplastic research, covering multiple dimensions from environmental distribution to ecological effects, and from the current pollution situation to risk assessment. In the future, microplastics research is expected to make breakthroughs in the following aspects: first, improving the efficient detection and quantitative methods of microplastics to accurately determine low-concentration microplastics in the environment; second, deeply exploring the combined pollution effects of microplastics with other pollutants (such as heavy metals and organic pollutants) and their long-term ecological impacts under complex environmental conditions; third, strengthening the source control and research and development of treatment technologies for microplastic pollution, and promoting the implementation of sustainable plastic management strategies to address the increasingly severe challenges of microplastic pollution. In conclusion, research into the impact of microplastics on soil and ecosystems must evolve in tandem with the changing times. By embracing more innovative approaches, aligning with global development trends, and fostering the green innovation of new materials and technologies, we can make significant strides in addressing this complex environmental challenge.

## 4. Discussion

### *4.1 The Development Process of the Research*

The standardized data of the sample research was imported into the CiteSpace software to draw the time zone view related to the research on the impact of MPs on soil and ecosystems abroad as shown in the figure 10. The figure clearly shows the evolution process of research hotspots and frontiers in this field, which is of great help for analyzing the development dynamics of the impact of MPs on soil and ecosystems abroad.

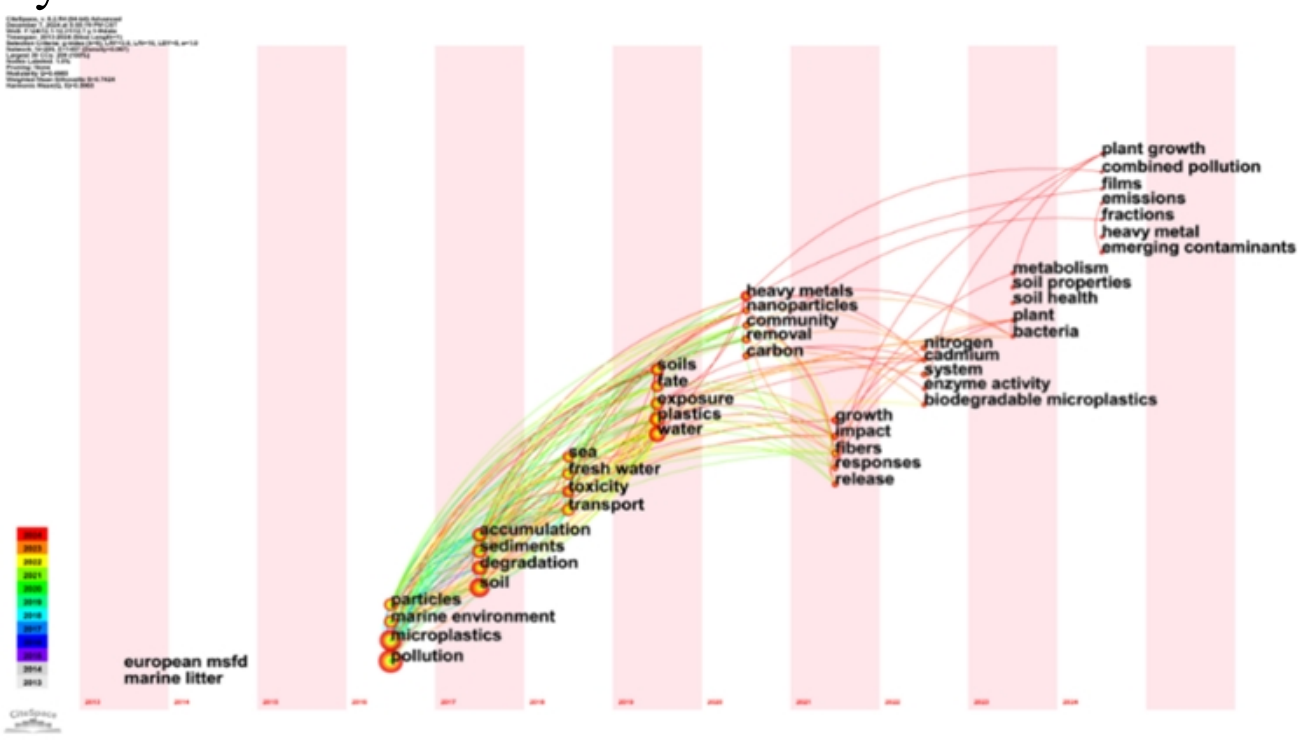

Figure. 10. A Time-Zone View o7f the Research Field on Soil Microplastic Pollution

From 2013 to 2016, it was the initial exploration stage. Researchers mainly focused on the distribution of microplastics in soil, pollution sources, and preliminary ecological impacts. Although the issue of microplastics has been of concern since the 1990s, the research on the impact of microplastics on the ecological environment began to increase around 2013, and the impact of microplastics on soil physical properties, plant growth, and soil microorganisms was also discussed [60]. From 2017 to 2020, it was the expansion stage. During this period, the research on the impact of microplastics on soil and ecosystems gradually deepened, and the research perspective became more diverse. During this period, the research on the negative impacts of microplastics on soil and ecosystems has gradually deepened, and the research perspectives have become more diversified, involving multiple disciplinary fields such as soil science, environmental science, ecology, biology, and chemistry For example, researchers began to explore the influence of the physical and chemical properties of different types of microplastics (such as polyethylene, polypropylene, polystyrene, etc.) on their behavior and ecological effects in soil [61]. At the same time, the research began to focus on the long-term accumulation effect of microplastics in soil and started to discuss the combined pollution effects of microplastics and other pollutants (such as heavy metals, pesticides, etc.) [62,63]. From 2021 to 2024, it was the integration stage. The research was no longer limited to the effects in a single environment but expanded to a comprehensive impact assessment of multiple environmental factors and biological communities. Scholars began to pay attention to the accumulation of microplastics in soil, the migration of microplastics in environmental media such as water and air, and the cross-media pollution effects of microplastics [64,65]. The research in this stage emphasized the interactive impact of microplastics on water and soil environments and began to pay attention to the application of policies and remediation technologies [66]. In the future, with the increasing attention to microplastics, future research will focus more on the multi-level assessment of the impact of microplastics on soil ecosystems and their effects on soil organisms. As research deepens, researchers will more clearly reveal the long-term hazards of microplastics to the ecological environment and propose practical solutions.

### *4.2 Analysis of Hotspots*

Based on the results of the aforementioned keyword analysis and in combination with the reading of specific literature, the current research hotspots in the field of soil microplastic pollution mainly include:

*4.2.1 Microplastic pollution in farmland soil.* The sources of microplastics in farmland soil are diverse. The study revealed that agricultural activities such as fertilization and irrigation are pathways for microplastics to enter farmland, and plastic film mulching is also an important source [67]. In modern agriculture, agricultural plastic films play a vital role, providing significant assistance in water retention and heat preservation for crops and enhancing crop yields. However, due to their lightweight and thin characteristics, the recycling of agricultural films is both difficult and costly, resulting in a large portion remaining in the soil, which gradually transform into microplastics in the environment. Microplastic pollution in farmland soil poses numerous hazards [68]. In the research by Long, they discovered that microplastics in farmland can alter the soil structure and properties, such as modifying the soil's bulk density and porosity, reducing soil fertility, and affecting crop growth [69]. Simultaneously, there is a risk of entering the human body and endangering health. According to the research by Huerta, L, his research team provided evidence for the first time that microplastics can be transferred through the soil-earthworm-chicken food chain, with the enrichment coefficient of microplastics in chicken manure reaching 105, confirming the existence of food chain transfer risks in microplastic pollution of farmland soil [70]. Thus, it is evident that microplastic pollution in farmland soil indeed has significant hazards and potential health and safety concerns.

*4.2.2 Bioremediation.* The second significant research direction is bioremediation; the two categories in the keyword clustering, "enzyme reaction activity" and "biodegradation", are both related to bioremediation. The extended words such as "soil microorganisms" and "biodegradable plastics" also appear in the keyword co-occurrence. Microplastics in the soil are complex organic substances with small sizes and stable chemical properties, and the effect of physicochemical remediation methods is poor. Many microplastics are highly resistant to biodegradation, and the degradation effect relying on conventional microorganisms in the soil is not satisfactory. Therefore, seeking more efficient microbial degradation methods for their degradation has become a research hotspot in the field of soil microplastic pollution. Simultaneously, the development of biodegradable plastic products to reduce soil microplastic pollution from the source has also attracted much attention [71]. Bioremediation utilizes native microorganisms in the soil or adds specific microorganisms to the soil to degrade pollutants in the soil. Current research has found that some insects, bacteria, and fungi in the soil have the ability to degrade microplastics. Cultivating microorganisms with high degradation efficiency is a research hotspot in the field of bioremediation. For instance, in the study by Hsiao, the researchers isolated two excellent fungi suitable for PBSA degradation from farmland, namely Aspergillus fumigatus L30 and Aspergillus terreus HC [72]. This study deepened the understanding of the degradation of PBSA films by filamentous fungi and provided insights into improving the biodegradation efficiency in the soil environment.

*4.2.3 Migration.* The migration of microplastics is also a current research hotspot; 6# migration in the migration keyword clustering, high-frequency keywords transport (259 times), feat (197 times), and the emergent word translocation in 2021-2022, all indicate that the migration process of microplastics in water, atmosphere, and soil is a research hotspot. In the study by Hoda, it was mentioned that microplastic pollution was first discovered in the marine environment, and then microplastics in the terrestrial environment were found [73]. Most plastic waste is generated and discharged on land, so there is migration of microplastics between the water and land environments. There is less research on the sources, destinations, and migration of microplastics in the terrestrial environment compared to that in the water environment [74]. However, studying the migration pathways of microplastics in the terrestrial environment can better determine the sources and destinations of microplastics in the soil, reduce soil microplastic pollution from the source, and decrease the possibility of secondary pollution caused by microplastics entering the groundwater system. Ren revealed the key processes of microplastic migration in the soil and groundwater ecosystems and explored the influencing factors of microplastic migration to the underground environment [75]. Currently, related research on the migration of soil microplastics is gradually receiving attention, and the number of studies is increasing, with a promising research prospect and space.

*4.2.4 Microplastic identification.* The high-frequency keyword "identification" appeared 300 times. Its appearance in the clustering indicates that the research on microplastic identification is an important research direction and hotspot in the field of soil microplastics. The identification and detection of microplastics are the prerequisites for discovering soil microplastic pollution and conducting pollution control. All research on soil microplastics is based on this, highlighting its importance. However, the detection of microplastics in the soil is highly challenging. The diameter of microplastics is less than 5 mm, and detection in the vast soil environment is difficult. The current identification methods mainly include Raman spectroscopy and Fourier transform infrared spectroscopy. However, these methods require complex sample collection and preprocessing, which is very time-consuming. Ai used hyperspectral imaging (HIS) technology to detect microplastics in the soil [76]. Hyperspectral imaging stands out in the field of soil microplastic analysis with the advantages of rapidity, non-destructiveness, and high efficiency, providing an efficient and non-destructive detection method. The specific principle is as follows: Due to differences in composition and functional groups, when a light source irradiates the surface of the object to be tested, different substances to be tested have different absorption rates, dispersion degrees, and reflectance for a certain wavelength. The similarity of spectral images of two samples indicates the similarity of their chemical composition and physical properties. Therefore, qualitative and quantitative analysis of the tested objects can be conducted by analyzing the differences between spectral signals. Compared with many spectral technologies, HSI simultaneously contains spectral information and spatial image information of the sample. Image information can be used to analyze the physical morphology characteristics of the target, and spectral information can be used to identify the chemical composition of the target. With the increasing research on microplastic identification and detection, it is believed that there will be more efficient and convenient identification and detection methods in the future.

*4.2.5 Impacts of soil microplastics on greenhouse gases.* So far, the research on soil microplastics has mainly focused on the impacts of microplastics on the biophysical properties of soil and the growth of crops [77-80]. However, in recent years, many studies have found that the pollution of MP in soil has exacerbated global warming, and the rise in global temperatures may further accelerate the aging and degradation of MP in the environment [81].It is known that soil-microcontaminant feedbacks are associated with the emissions of three major greenhouse gases: $CO_2$, nitrous oxide $N_2O$, and methane $CH_4$ [82-84]. Microplastics pollution can lead to 1–32% increases in $CO_2$ emissions, 12–50% increases in $N_2O$ emissions, and 14–60% changes in $CH_4$ emissions from soils [85,86]. The presence of soil microplastics may promote the decomposition of soil organic matter, thereby increasing $CO_2$ emissions. As a primary driver of climate change, sustained increases in $CO_2$ emissions will lead to irreversible climate changes that further threaten human societies [87]. Changes in $CO_2$ emissions from agricultural soils due to microplastics are closely related to microplastics types, concentrations, soil conditions, and biological activities. In scenarios with only microplastics additions, 10% low-density polyethylene (LDPE) significantly reduced $CO_2$ emission fluxes by 26.5–33.9% in wet and red soils, whereas 18% LDPE increased $CO_2$ emission fluxes by 28.67% in agricultural soils by altering soil properties and microbial communities [88,89]. The effects of microplastics on cumulative $CO_2$ emissions are dose-dependent: high-dose microplastics (1.00%) promote soil $CO_2$ emissions, while low-dose microplastics (0.01% and 0.10%) have negligible effects [90]. This is because high-dose microplastics can enhance soil microbial growth and activity by accelerating microbial metabolism, increasing dissolved organic carbon (DOC) content, or elevating oxygen concentrations for microbial respiration [91,92]. In addition to microplastic types and concentrations, the degree of microplastics aging is a critical factor, as aging often induces significant changes in microplastics morphology and surface functional groups, altering their physicochemical properties. Compared with unaged microplastics, aged microplastics increase $CO_2$ emissions by improving microbial metabolic behavior, which is associated with their higher oxygen-containing functional groups and adsorption areas [93]. For $N_2O$, its production in soils is primarily linked to nitrification and denitrification processes in the nitrogen cycle. The presence of microplastics may disrupt the structure and function of microbial communities involved in these processes [94]. Some studies have found that microplastics additions alter the abundance and activity of key microorganisms such as ammonia-oxidizing bacteria (AOB), ammonia-oxidizing archaea (AOA), and denitrifying bacteria. Such changes in microbial communities can disrupt the balance between nitrification and denitrification, affecting $N_2O$ production and emissions [95]. Inhibited nitrification with relatively enhanced denitrification may increase $N_2O$ emissions, whereas inhibited denitrification may reduce them. Field experiments have also shown that applying urea alongside microplastics significantly alters soil $N_2O$ emission fluxes [96]. Certain microplastics may stimulate the growth and metabolism of methanogens, increasing $CH_4$ production. For example, experiments with polyethylene MP additions detected significantly increased $CH_4$ emissions in soils [97]. This may occur because microplastic provide additional carbon sources for methanogens or alter soil microenvironments to favor their survival and reproduction [98]. Moreover, microplastics may promote the decomposition of soil organic matter, providing more substrates for methanogens. Microplastics effects on soil water retention and distribution also play roles: excessively high soil moisture enhances soil anaerobiosis, favoring $CH_4$ production; conversely, low moisture may inhibit the activity of methanogens and methane-oxidizing bacteria, reducing $CH_4$ production and consumption. Current research on soil microplastics and greenhouse gas emissions remains limited, with their impacts marked by complexity and uncertainty. microplastics differentially affect $CO_2$, $N_2O$, and $CH_4$ via altering soil properties, microbial communities, and biogeochemical cycles. Future studies require multidisciplinary approaches integrating long-term observations, microbiomics, and modeling to fully assess microplastics pollution’s role in climate change and inform sustainable agriculture.

## 5.Conclusions

In this research, we utilized CiteSpace software to analyze a considerable number of literatures and conducted a visual examination of the academic progress and achievements in this field. We summarized the research hotspots and proposed the key points and directions for future research in the domain of soil microplastics. As can be clearly seen from the publication trend graph, the overall research on the impact of microplastics on soil and ecosystems is on the rise. This suggests that scholars are increasingly devoting themselves to this area of study. With the growing environmental awareness among the public and extensive media coverage, it is foreseeable that people's attention to microplastic pollution will soar in the future.In terms of the authors of the literature and their collaborative relationships, the above analysis reveals a significant disparity in the publication levels of those researching the impact of microplastics on soil and ecosystems. Only a small number of researchers have delved deeply into this topic from multiple perspectives. In contrast, the majority have merely conducted single - time studies on a particular aspect, failing to carry out systematic and comprehensive investigations. Evidently, the investment in this field remains inadequate.Judging from the number of papers published by research institutions, some of them have emerged as the leading forces in international research concerning the impact of microplastics on soil and ecosystems. These institutions are engaged in extensive transnational collaborations and are geographically dispersed.By looking at the research on the impact of microplastics on soil and ecosystems in the database, it is evident that countries with a high volume of publications basically initiated their studies in this field simultaneously. This implies that the technological development in this regard is largely synchronized across different nations. In the face of the worsening plastic pollution issue, the Chinese government has been gradually ramping up its focus on plastic pollution control. From the cluster analysis perspective, the current stage of research exhibits characteristics such as clustering, diversification, and rapid growth. Based on various themes, scholars have spawned a multitude of studies exploring the impact of microplastics on soil and ecosystems. In conclusion, the research achievements in the field of soil microplastic pollution have helped people recognize the hazards of microplastics. However, to maintain soil health and precisely block the intake of microplastics by organisms, more detailed pollution assessment and treatment methods are still required.

## Acknowledgements

This study was funded by the National Natural Science Foundation of China (51968073, 32460317) and the Natural Science Foundation of Jilin Province Department of Science and Technology (20210101089JC, YDZJ202501ZYTS403).

## Conflict of interest

The authors declare no competing interests.

## References

[1]Hernandez E, Nowack B and Mitrano D M 2017 Polyester textiles as a source of microplastics from households: a mechanistic study to understand microfiber release during washing *Environ. Sci. Technol.* **51** 7036-7046.
[2]Thompson R C, Olsen Y, Mitchell R P, Davis A, Rowland S J, John-Anthony W G, McGonigle D and Russell A E 2004 Lost at sea: where is all the plastic? *Science*. **304** 838-838.
[3]Rios L M, Moore C and Jones P R 2007 Persistent organic pollutants carried by synthetic polymers in the ocean environment *Mar. Pollut. Bull.* **54** 1230-1237.
[4]Gouin T, Roche N, Lohmann R and Hodges G 2011 A thermodynamic approach for assessing the environmental exposure of chemicals absorbed to microplastic *Environ. Sci. Technol.* **45** 1466-1472.
[5]McCormick A, Hoellein T J, Mason S A, Schlueo J and Kelly J J 2014 Microplastic is an abundant and distinct microbial habitat in an urban river *Environ. Sci. Technol.* **48** 11863-11871.
[6]Lusher A L, Burke A, O'Connor I and Officer R 2014 Microplastic pollution in the Northeast Atlantic Ocean: validated and opportunistic sampling *Mar. Pollut. Bull.* **88** 325-333.
[7]Lebreton L and Andrady A 2019 Future scenarios of global plastic waste generation and disposal *Palgrave. Communications*. **5** 1-11.
[8]De Souza Machado A A, Kloas W, Zarfl C, Hempel S and Rillig M 2018 Microplastics as an emerging threat to terrestrial ecosystems *Global. Change. Biol*. **24** 1405-1416.
[9]Astner A F, Hayes D G, O'Neill H, Evans B R, Pingali S V, Urban V S and Young T M 2019 Mechanical formation of micro-and nano-plastic materials for environmental studies in agricultural ecosystems *Sci. Total. Environ*. **685** 1097-1106.
[10]Han X, Lu X and Vogt R D 2019 An optimized density-based approach for extracting microplastics from soil and sediment samples *Environ. Pollut*. **254** 113009.
[11]Leslie H A, Brandsma S H, Van Velzen M J M and Vethaak A D 2017 Microplastics en route: Field measurements in the Dutch river delta and Amsterdam canals, wastewater treatment plants, North Sea sediments and biota *Environ. Int.* **101** 133-142.
[12]Bläsing M and Amelung W 2018 Plastics in soil: Analytical methods and possible sources *Sci. Total. Environ*. **612** 422-435.
[13]Su Y L, Zhang Z J, Wu D, Zhan L, Shi H H and Xie B 2019 Occurrence of microplastics in landfill systems and their fate with landfill age *Water. Res*. **164** 114968.
[14]Yu J R, Adingo S, Liu X L, Li X D, Sun J and Zhang X N 2022 Micro plastics in soil ecosystem-A review of sources, fate, and ecological impact *Plant. Soil. Environ*. **68** 1-17.
[15]Gui X Y, Ren Z F, Xu X Y, Chen X, Chen M, Wei Y Q, Zhao L, Qiu H, Gao B and Cao X D 2022 Dispersion and transport of microplastics in three water-saturated coastal soils *J. Hazard. Mater*. **424** 127614.
[16]Rillig M C, Leifheit E and Lehmann J 2021 Microplastic effects on carbon cycling processes in soils *Plos. Biol*, **19** e3001130.
[17]Mbachu O, Jenkins G, Kaparaju P and Pratt C 2021 The rise of artificial soil carbon inputs: Reviewing microplastic pollution effects in the soil environment *Sci. Total. Environ*. **780** 146569.
[18]Mato Y, Isobe T, Takada H, Kanehiro H, Ohtake C and Kaminuma T 2001 Plastic resin pellets as a transport medium for toxic chemicals in the marine environment *Environ. Sci. Technol.* **35** 318-324.
[19]Zhang F, Zhao Y T, Wang D D, Yan M Q, Zhang J, Zhang P Y, Ding T G, Chen L and Chen C 2021 Current technologies for plastic waste treatment: A review *J. Clean. Prod*. **282** 124523.
[20]Alimba C G and Faggio C 2019 Microplastics in the marine environment: Current trends in environmental pollution and mechanisms of toxicological profile *Environ. Toxicol. Phar.* **68** 61-74.
[21]Abidli S, Pinheiro M, Lahbib Y, Neuparth T, Santos M M and Trigui El Menif N 2021 Effects of environmentally relevant levels of polyethylene microplastic on Mytilus galloprovincialis (Mollusca: Bivalvia): filtration rate and oxidative stress *Environ. Sci. Pollut. R*. **28** 26643-26652.
[22]Ouyang M Y, Liu J H, Wen B, Huang J N, Feng X S, Gao J Z and Chen Z Z 2021 Ecological stoichiometric and stable isotopic responses to microplastics are modified by food conditions in koi carp *J. Hazard. Mater*. **404** 124121.
[23]Wang Q L, Adams C A, Wang F Y, Sun Y H and Zhang S W 2022 Interactions between microplastics and soil fauna: a critical review *Crit. Rev. Env. Sci. Tec.* **52** 3211-3243.
[24]Chai B W, Wei Q, She Y Z, Lu G N, Dang Z and Yin H 2020 Soil microplastic pollution in an e-waste dismantling zone of China *Waste. Manage.* **118** 291-301.

[25]Pérez-Reverón R, Álvarez Méndez S J, Kropp R M, Perdomo González A, Hernández Borges J and Díaz-Peña F 2022 Microplastics in agricultural systems: analytical methodologies and effects on soil quality and crop yield *Agriculture.* **12** 1162.
[26]Jin T Y, Tang J C, Lyu H H, Wang L, Gillmore A B and Schaeffer S M 2022 Activities of microplastics (MPs) in agricultural soil: a review of MPs pollution from the perspective of agricultural ecosystems *J. Agric. Food Chem.* **70** 4182-4201.
[27]Yu H, Zhang Y, Tan W B and Zhang Z 2022 Microplastics as an emerging environmental pollutant in agricultural soils: effects on ecosystems and human health *Front. Env. SCI-SWITZ.* **10** 855292.
[28]Horton A A, Walton A, Spurgeon D J, Lahive E and Svendsen C 2017 Microplastics in freshwater and terrestrial environments: Evaluating the current understanding to identify the knowledge gaps and future research priorities *Sci. Total. Environ*. **586** 127-141.
[29]De Souza Machado A A, Lau C W, Kloas W, Bergmann J, Bachelier J B, Faltin E, Becker R, Görlich A S and Rilling M C 2019 Microplastics can change soil properties and affect plant performance *Environ. Sci. Technol.* **53** 6044-6052.
[30]De Souza Machado A A, Lau C W, Till J, Kloas W, Lehmann A, Becker R 2018 Rilling M 2018 Impacts of microplastics on the soil biophysical environment *Environ. Sci. Technol.* **52** 9656-9665.
[31]Zhang G S 2018 Liu Y F 2018 The distribution of microplastics in soil aggregate fractions in southwestern China *Sci. Total. Environ*. **642** 12-20.
[32]Liu M T, Lu S B, Song Y, Lei L L, Hu J N, Lv W W, Zhou W Z, Cao C J, Shi H H, Yang X F and He D F 2018 Microplastic and mesoplastic pollution in farmland soils in suburbs of Shanghai, China *Environ. Pollut*. **242** 855-862.
[33]Scheurer M and Bigalke M 2018 Microplastics in Swiss floodplain soils *Environ. Sci. Technol*. **52** 3591-3598.
[34]Boots B, Russell C W and Green D S 2019 Effects of microplastics in soil ecosystems: above and below ground *Environ. Sci. Technol*. **53** 11496-11506.
[35]Qi Y L, Yang X M, Pelaez A M, Lwanga E H, Beriot N, Gertsen H, Garbeva P and Geissen V 2018 Macro-and micro-plastics in soil-plant system: effects of plastic mulch film residues on wheat (Triticum aestivum) growth *Sci. Total. Environ* **645** 1048-1056.
[36]Guo J J, Huang X P, Xang L, Wang Y Z, Li Y W, Li H, Cai Q Y, Mo C H and Wong M H 2020 Source, migration and toxicology of microplastics in soil *Environ. Int*. **137** 105263.
[37]He D F, Luo Y M, Lu S B, Liu M T, Song Y and Lei L L 2018 Microplastics in soils: Analytical methods, pollution characteristics and ecological risks *Trac-Trend. Anal. Chem*. **109** 163-172.
[38]Corradini F, Meza P, Eguiluz R, Casado F, Lwanga E H and Geissen V 2019 Evidence of microplastic accumulation in agricultural soils from sewage sludge disposal *Sci. Total. Environ*, **671** 411-420.
[39]Lwanga E H, Beriot N, Corradini F, Silva V, Yang X M, Baartman J, Rezaei M, van Schaik L, Riksen M and Geissen V 2022 Review of microplastic sources, transport pathways and correlations with other soil stressors: a journey from agricultural sites into the environment *Chem. Biol. Technol. A.* **9** 20.
[40]Rezaei M, Riksen M J P M, Sirjani E, Sameni A and Geissen V 2019 Wind erosion as a driver for transport of light density microplastics *Sci. Total. Environ*. **669** 273-281.
[41]Yu Y, Chen Y H, Wang Y, Xue S, Liu M J, Tang D W S, Yang X M and Geissen V 2023 Response of soybean and maize roots and soil enzyme activities to biodegradable microplastics contaminated soil *Ecotox. Environ. Safe*. **262** 115129.
[42]Ju H, Yang X M, Osman R and Geissen V 2023 Effects of microplastics and chlorpyrifos on earthworms (Lumbricus terrestris) and their biogenic transport in sandy soil. *Environ. Pollut* **316** 120483.
[43]Meng F R, Yang X M, Riksen M, Xu M G and Geissen V 2021 Response of common bean (Phaseolus vulgaris L.) growth to soil contaminated with microplastics *Sci. Total. Environ*. **755** 142516.
[44]Corradini F, Bartholomeus H, Lwanga E H, Gertsen H and Geissen V 2019 Predicting soil microplastic concentration using vis-NIR spectroscopy *Sci. Total. Environ*. **650** 922-932.
[45]Zhang S L, Yang X M, Gertsen H, Peters P, Salánki T and Geissen V 2018 A simple method for the extraction and identification of light density microplastics from soil *Sci. Total. Environ*. **616** 1056-1065.
[46]Qi R M, Jones D L, Li Z, Liu Q and Yan C R 2020 Behavior of microplastics and plastic film residues in the soil environment: A critical review *Sci. Total. Environ*. **703** 134722.
[47]Wei S W, Zhao Y J, Zhou R M, Lin J W, Su T T, Tong H B and Wang Z Y 2022 Biodegradation of polybutylene adipate-co-terephthalate by Priestia megaterium, Pseudomonas mendocina, and Pseudomonas pseudoalcaligenes following incubation in the soil *Chemosphere.* **307** 135700.
[48]Liu H F, Yang X M, Liu G B, Liang C T, Xue S, Chen H, Ritsema C J and Geissen V 2017 Response of soil dissolved organic matter to microplastic addition in Chinese loess soil *Chemosphere.* **185** 907-917.
[49]Qiu X R, Ma S R, Pan J R, Cui Q, Zheng W, Ding L, Liang X J, Xu B L, Guo X T and Rilling M C 2024 Microbial metabolism influences microplastic perturbation of dissolved organic matter in agricultural soils *ISME. J.* **18** wrad017.
[50]Royer S J, Greco F, Kogler M and Deheyn D D 2023 Not so biodegradable: Polylactic acid and cellulose/plastic blend textiles lack fast biodegradation in marine waters *PLoS. One*. **18** e0284681.
[51]Fang C, He Y L, Yang Y T, Fu B, Pan S T, Jiao F, Wang J and Yang H R 2023 Laboratory tidal microcosm deciphers responses of sediment archaeal and bacterial communities to microplastic exposure *J. Hazard. Mater.* **458** 131813.
[52]Zhou J L, Chen M Y, Li Y, Wang J J, Chen G L and Wang J 2024 Microbial bioremediation techniques of microplastics and nanoplastics in the marine environment *Trac-Trend. Anal. Chem*. 117971.
[53]Ranauda M A, Tartaglia M, Zuzolo D, Prigioniero A, Maisto M, Fosso E, Sciarrillo R and Guarino, C 2024 From the rhizosphere to plant fitness: Implications of microplastics soil pollution *Environ. Exp. Bot*. **226** 105874.
[54]Fan C Z, Li Y X, Tian C Q and Li Z Y 2024 Effects of microplastics on soil C and N cycling with or without interactions with soil amendments or soil fauna *Eur. J. Soil. Sci.* **75** e13446.

[55]Tian L L, Cheng J J, Ji R, Ma Y N and Yu X Y 2022 Microplastics in agricultural soils: sources, effects, and their fate *Curr. Opin. Env. Sci. Hl.* **25** 100311.
[56]Thompson R C, Courtene Jones W, Boucher J, Pahl S, Raubenheimer K and Koelmans A A 2024 Twenty years of microplastic pollution research—what have we learned? *Science*. **386** eadl2746.
[57]Zheng B H, Zhou L, Wang J N, Dong, P C, Zhao T, Deng Y T, Song L R, Shi J Q and Wu Z X 2025 The shifts in microbial interactions and gene expression caused by temperature and nutrient loading influence Raph idiopsis raciborskii blooms *Water. Res*. **268** 122725.
[58]Löhr A, Savelli H, Beunen R, Kalz M, Ragas A and Belleghem F V 2017 Solutions for global marine litter pollution *Curr. Opin. Env. Sust*. **28** 90-99.
[59]Wilcox C, Mallos N J, Leonard G H, Rodriguez A and Hardesty B D 2016 Using expert elicitation to estimate the impacts of plastic pollution on marine wildlife *Mar. Policy*, **65** 107-114.
[60]Ding L, Huang D F, Ouyang Z Z and Guo X T 2022 The effects of microplastics on soil ecosystem: A review *Curr. Opin. Env. Sci. Hl.* **26** 100344.
[61]Guo W J, Ye Z W, Zhao Y N, Lu Q L, Shen B, Zhang X, Zhang W F, Chen S C, Li Y 2024 Effects of different microplastic types on soil physicochemical properties, enzyme activities, and bacterial communities *Ecotox. Environ. Safe*. **286** 117219.
[62]Awet T T, Kohl Y, Meier F, Straskraba S, Grün A L, Ruf T, Jost C, Drexel R, Tunc E and Emmerling, C 2018 Effects of polystyrene nanoparticles on the microbiota and functional diversity of enzymes in soil *Environ. Sci. Eur.* **30** 1-10.
[63]Bosker T, Bouwman L J, Brun N R, Behrens P and Vijver M G 2019 Microplastics accumulate on pores in seed capsule and delay germination and root growth of the terrestrial vascular plant Lepidium sativum *Chemosphere.* **226** 774-781.
[64]Zhao S L, Zhang Z Q, Chen L, Cui Q L, Cui Y X, Song D X and Fang L C 2022 Review on migration, transformation and ecological impacts of microplastics in soil *Appl. Soil. Ecol.* **176** 104486.
[65]Singh S S, Chanda R, Singh N S, Ramtharmawi, Devi N R, Devi K V, Upadhyay K K and Tripathi S K 2024 Microplastic pollution: exploring trophic transfer pathways and ecological impacts *Discover Environment*. **2** 103.
[66]Wei J, Chen M and Wang J 2023 Insight into combined pollution of antibiotics and microplastics in aquatic and soil environment: Environmental behavior, interaction mechanism and associated impact of resistant genes *Trac-Trend. Anal. Chem*. **166** 117214.
[67]Hu J, Zhang L Q, Zhang W Y, Muhammad I, Yin C Y, Zhu Y X, Li C and Zheng LG 2024 Significant influence of land use types and anthropogenic activities on the distribution of microplastics in soil: A case from a typical mining-agricultural city *J. Hazard. Mater.* **477** 135253.
[68]Zhou J, Jia R, Brown R W, Yang Y D, Zeng Z H, Jones D L and Zang H D 2023 The long-term uncertainty of biodegradable mulch film residues and associated microplastics pollution on plant-soil health *J. Hazard. Mater.* **442** 130055.
[69]Long B B, Li F Y, Wang K, Huang Y Z, Yang Y J and Xie D 2023 Impact of plastic film mulching on microplastic in farmland soils in Guangdong province, China *Heliyon*. **9**.
[70]Huerta Lwanga E, Mendoza Vega J, Ku Quej V, Cid J D L A, Cid L S D, Chi C, Segura G S, Gertsen H, Salánki T, Ploeg M V D, Koelmans A A and Geissen V 2017 Field evidence for transfer of plastic debris along a terrestrial food chain *Sci, Rep-UK*. **7** 14071.
[71]Song T J, Liu J X, Han S Q, Li Y, Xu T Q, Xi J, Hou L J and Lin Y B 2024 Effect of conventional and biodegradable microplastics on the soil-soybean system: A perspective on rhizosphere microbial community and soil element cycling *Environ. Int*. **190** 108781.
[72]Chien H L, Tsai Y T, Tseng W S, Wu J A, Kuo S L, Chang S L, Huang S J and Liu C T 2022 Biodegradation of PBSA films by Elite Aspergillus isolates and farmland soil *Polymers*, **14** 1320.
[73]Fakour H, Lo S L, Yoashi N T, Massao A M, Lema N N, Mkhontfo F B, Jomalema P C, Jumanne N S, Mbuya B H, Mtweve J T and Imani M 2021 Quantification and analysis of microplastics in farmland soils: characterization, sources, and pathways *Agriculture*. **11** 330.
[74]Zhang S L, Wang W, Yan P K, Wang J Q, Yan S H, Liu X B and Aurangzeib M 2023 Microplastic migration and distribution in the terrestrial and aquatic environments: A threat to biotic safety *J. Environ. Manage*. **333** 117412.
[75]Ren Z F, Gui X Y, Xu X Y, Zhao L, Qiu H and Cao X D 2021 Microplastics in the soil-groundwater environment: aging, migration, and co-transport of contaminants–a critical review *J. Hazard. Mater.* **419** 126455.
[76]Ai W J, Liu S L, Liao H P, Du J Q, Cai Y L, Liao C L, Shi H W, Lin Y D, Junaid M, Yue X J and Wang J 2022 Application of hyperspectral imaging technology in the rapid identification of microplastics in farmland soil *Sci. Total. Environ*. **807** 151030.
[77]de Souza Machado A A, Lau C W, Kloas W, Bergmann J, Bachelier J B, Faltin E, Becker R, Gorlich A S and Rillig M C 2019. Microplastics can change soil properties and affect plant performance *Environ. Sci. Technol.* **53** 6044–6052.
[78]Iqbal S, Xu J, Allen S D, Khan S, Nadir S, Arif M S and Yasmeen T 2020. Unraveling consequences of soil micro-and nano-plastic pollution on soil-plant system: implications for nitrogen (N) cycling and soil microbial activity *Chemosphere*. **260** 127578.
[79]Iqbal S, Xu J, Khan S, Arif M S, Yasmeen T, Nadir S and Schaefer D A 2021. Deciphering microplastic ecotoxicology: impacts on crops and soil ecosystem functions *Circular Agricultural Systems* **1** 1–7.
[80]Wang F, Wang Q, Adams C A, Sun Y and Zhang S 2022. Effects of microplastics on soil properties: current knowledge and future perspectives *J. Hazard Mater.* **424** 127531.
[81]Li K, Du L, Qin C, Bolan N, Wang H and Wang H 2024 Microplastic pollution as an environmental risk exacerbating the greenhouse effect and climate change: a review *Carbon Res.* **3** 9.
[82]Gao B, Li Y, Zheng N, Liu C, Ren H and Yao H 2022 Interactive effects of microplastics, biochar, and earthworms on CO2 and N2O emissions and microbial functional genes in vegetable-growing soil *Environ. Res.* **213** 113728.

[83]Gao B, Yao H, Li Y and Zhu Y 2021 Microplastic addition alters the microbial community structure and stimulates soil carbon dioxide emissions in vegetablegrowing soil *Environ. Toxicol. Chem.* **40** 352–365.
[84]Xiao M, Shahbaz M, Liang Y, Yang J, Wang S, Chadwicka D R, Jones D, Chen J and Ge T 2021 Effect of microplastics on organic matter decomposition in paddy soil amended with crop residues and labile C: a three-source-partitioning study *J. Hazard Mater.* **416** 126221.
[85]Han L, Zhang B, Li D, Chen L, Feng Y, Xue L, He J and Feng Y 2022 Co-occurrence of microplastics and hydrochar stimulated the methane emission but suppressed nitrous oxide emission from a rice paddy soil *J. Clean. Prod.* **337** 130504.
[86]Inubushi K, Kakiuchi Y, Suzuki C, Sato M, Ushiwata S Y and Matsushima M Y 2022 Effects of biodegradable plastics on soil properties and greenhouse gas production *Soil Sci. Plant Nutr.* **68** 183–188.
[87]Wei N and Xia J Y 2024. Robust projections of increasing land carbon storage in boreal and temperate forests under future climate change scenarios *One Earth* **7 (1)** 88–99.
[88]Gao B, Yao H and Li Y, Zhu Y 2021 Microplastic addition alters the microbial community structure and stimulates soil carbon dioxide emissions in vegetablegrowing soil *Environ. Toxicol. Chem.* **40 (2)** 352–365.
[89]Yu H, Zhang Z, Zhang Y, Song Q, Fan P, Xi B and Tan W 2021 Effects of microplastics on soil organic carbon and greenhouse gas emissions in the context of straw incorporation: a comparison with different types of soil *Environ. Pollut.* **288** 117733.
[90]Zhang Y, Li X, Xiao M, Feng Z, Yu Y and Yao H 2022. Effects of microplastics on soil carbon dioxide emissions and the microbial functional genes involved in organic carbon decomposition in agricultural soil *Sci. Total Environ.* **806 (Pt 3)** 150714.
[91]Gao B, Yao H, Li Y and Zhu Y 2021 Microplastic addition alters the microbial community structure and stimulates soil carbon dioxide emissions in vegetablegrowing soil *Environ. Toxicol. Chem.* **40** 352–365.
[92]Liu H, Yang X, Liu G, Liang C, Xue S, Chen H, Ritsema C J and Geissen V 2017 Response of soil dissolved organic matter to microplastic addition in Chinese loess soil *Chemosphere* **185** 907–917.
[93]Chen M L, Liu S S, Bi M H, Yang X Y and Deng R Y 2022 Aging behavior of microplastics affected DOM in riparian sediments: from the characteristics to bioavailability *J. Hazard. Mster.* **431** 128522.
[94]Seeley M E, Song B, Passie R and Hale R C. 2020 Microplastics affect sedimentary microbial communities and nitrogen cycling *Nat. Commun.* **11** 12372.
[95]Huang S Y, Guo T, Feng Z, Li B C, Cai Y M, Ouyang D, Gustave W, Ying C F and Zhang, H B 2023 Polyethylene and polyvinyl chloride microplastics promote soil nitrification and alter the composition of key nitrogen functional bacterial groups *J. Hazard. Mster.,* **453** 131391.
[96]Rillig M C, Hoffmann M, Lehmann A, Lück M and Augustin Jürgen 2021 Microplastic fibers affect dynamics and intensity of CO 2 and N 2 O fluxes from soil differently. *Microplastics and Nanoplastics* **1** 1-11.
[97]Zhang Z H, Yang Z H, Yue H W, Xiao M, Ge T D, Li Y Y, Yu Y X and Yao H Y 2023 Discrepant impact of polyethylene microplastics on methane emissions from different paddy soils *Appl. Soil. Ecol.* **181** 104650.
[98]Han L, Zhang B, Li D, Chen L, Feng Y, Xue L, He J and Feng Y 2022. Co-occurrence of microplastics and hydrochar stimulated the methane emission but suppressed nitrous oxide emission from a rice paddy soil *J. Clean. Prod.* **337** 130504.